%% file: lyu_azarpeyvand_sinayoko_2015v1.0.tex
\pdfoutput=1
\documentclass{jfm}

\usepackage{graphicx}
\usepackage{epstopdf}
\usepackage{natbib}

\input{jfmdefs.tex}

\usepackage{nomencl}%   nomenclature generation via makeindex
\makenomenclature
\usepackage{subfigure}% subcaptions for subfigures
\usepackage{amsmath}
\usepackage{bm}
\usepackage{layouts}

\newcommand{\ud}{\,\text{d}}
\title{Prediction of noise from serrated trailing-edges}
\author[B. Lyu, M. Azarpeyvand and S. Sinayoko]%
{B. Lyu$^1$,\ns
M. Azarpeyvand$^2$
  \thanks{Email address for correspondence: m.azarpeyvand@bristol.ac.uk}\break
and S. Sinayoko$^3$}

\affiliation{$^1$Department of Engineering, University of Cambridge, Cambridge CB2 1PZ, UK\\[\affilskip]
$^2$Department of Mechanical Engineering, University of Bristol, Bristol BS8 1TR, UK\\[\affilskip]
$^3$Institute of Sound and Vibration Research, University of Southampton, Sounthampton SO17 1BJ, UK}

\begin{document}
\maketitle
\begin{abstract}
  A new analytical model is developed for the prediction of noise from serrated trailing-edges. The model generalizes Amiet's trailing-edge noise theory to sawtooth trailing-edges, resulting in an inhomogeneous partial differential equation. The equation is then solved by means of a Fourier expansion technique combined with an iterative procedure. The solution is validated through comparison with finite element method for a variety of serrations at different Mach numbers. Results obtained using the new model predict noise reduction of up to 10 dB at 90$^{\circ}$ above the trailing-edge, which is more realistic than predictions based on Howe's model and also more consistent with experimental observations. A thorough analytical and numerical analysis of the physical mechanism is carried out and suggests that the noise reduction due to serration originates primarily from interference effects near the trailing-edge. A closer inspection of the proposed mathematical model has led to the development of two criteria for the effectiveness of the trailing-edge serrations, consistent but more general than those proposed by Howe. While experimental investigations often focus on noise reduction at ninety degrees above the trailing-edge, the new analytical model shows that the destructive interference scattering effects due to the serrations cause significant noise reduction at large polar angles, near the leading edge. It has also been observed that serrations can significantly change the directivity characteristics of the aerofoil at high frequencies and even lead to noise increase at high Mach numbers.
\end{abstract}
\printnomenclature% creates nomenclature section produced by MakeIndex
\section{Introduction}
%\begin{figure}
%    \centering
%    \includegraphics[width=\figurewidth]{TrailingEdgeScattering.pdf}
%	\caption{Schematic illustration of the trailing-edge noise}
%	\label{fig:TrailingEdgeScattering}
%\end{figure}
The past few decades has seen a rapid growth of air traffic, while the public's attention to aircraft noise and its health consequences has also been continuously increasing. This has led to more stringent regulations for aircraft noise~\citep{Casalino2008}. With regard to the impact of aircraft on community noise, the take-off and landing process are of main concern. Among the different mechanisms present during the landing process, airframe noise is believed to be the dominant component. It is widely accepted that the broadband noise, induced by the interaction of boundary layer with the aerofoil trailing-edge, known as the turbulent boundary layer trailing-edge noise, plays a significant role in the overall airframe noise.
Turbulent boundary layer trailing-edge noise also dominates the noise produced by wind turbines~\citep{Oerlemans2007}. Unless explicitly stated, the turbulent boundary layer trailing-edge noise will be referred to as trailing-edge noise in the rest of this paper.

When a turbulent boundary layer convects past the trailing-edge, unsteady pressure with a wavenumber in the hydrodynamic range is scattered into sound~\citep{Chase1975}. Both experiments and theory reveal that the radiated sound power varies with the flow velocity to the power of 5, which is more efficient, at low Mach numbers, compared to the power of 8 valid for free stream flows~\citep{Lighthill1952, Williams1970}.
\begin{figure}
    \centering
    \includegraphics[width=0.7\textwidth]{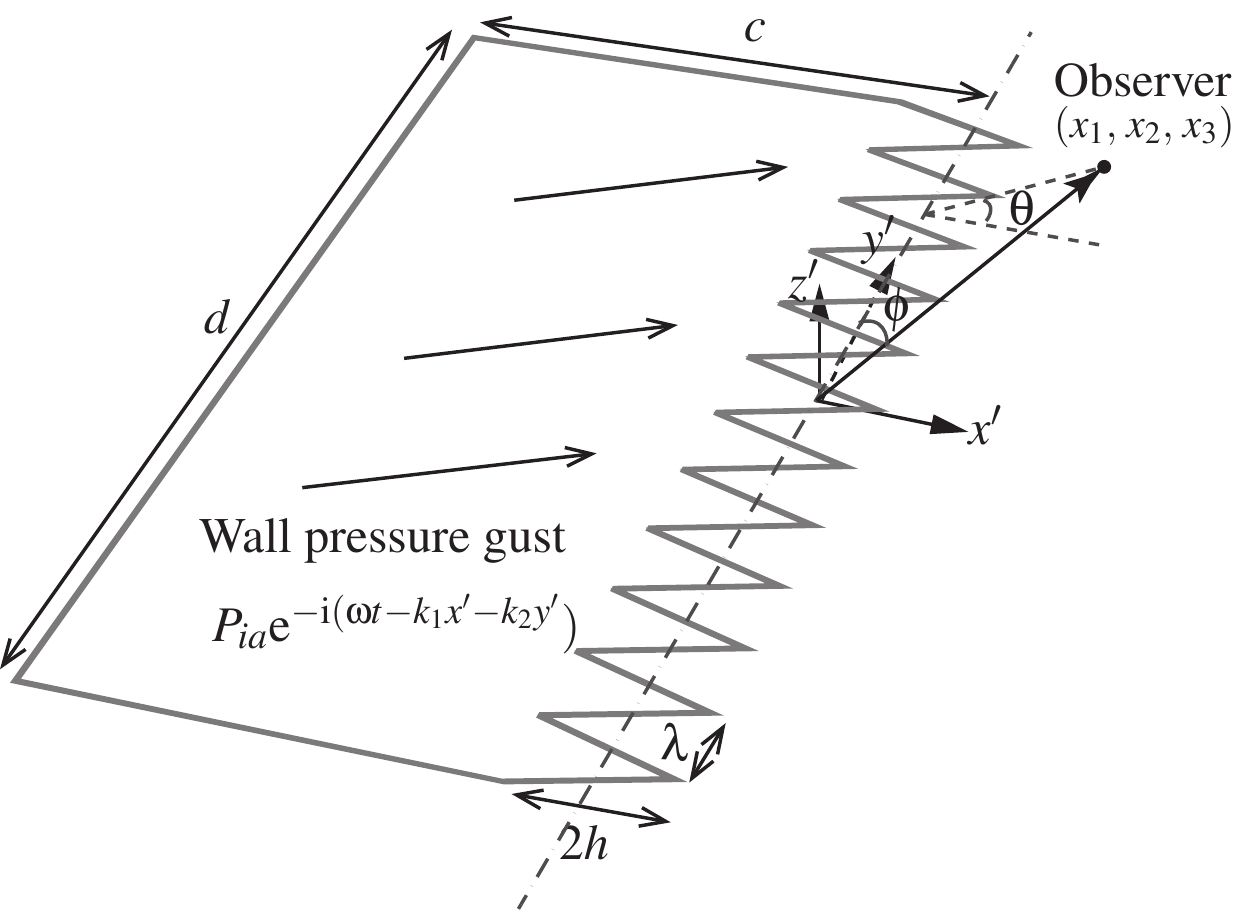}
    \label{fig:SerratedFlatPlate}
    \caption{The schematic of a flat plate with and trailing-edge serrations.}
\end{figure}

Different models have been put forward for predicting and understanding trailing-edge noise. In 1976, \citet{Amiet1976, Amiet1978} proposed a semi-analytical model in which the aerofoil is modelled as a flat plate. The model followed Schwarzschild's technique to obtain the scattered pressure on the surface of the flat plate and the far-field sound was obtained using the surface pressure integral based on the theories of Kirchoff and Curle. The model established an analytical relationship between the far-field sound spectral density and the wavenumber spectral density of the wall surface pressure under the turbulent boundary layer. Amiet's model agrees well with experimental observations, especially at high frequencies~\citep{Roger2005}. It is worth noting that Amiet's model assumed that the leading edge of the plate is infinitely far away from the trailing-edge and thus has no effects on the scattered pressure, which might not be accurate at low frequencies. In 2005, in order to investigate the leading-edge back-scattering effects, \citet{Roger2005} extended Amiet's model by incorporating the back-scattered pressure from the leading edge and found that when the Helmholtz number $kc > 1$, the back-scattering can be safely ignored and only at very low frequencies does the back-scattering alter the far-field sound.

As trailing-edge noise dominates the sound generation at low Mach numbers, different noise reduction techniques have been investigated. Howe proposed a theoretical model to predict the sound generated by a semi-infinite plate with serrated trailing-edge of sinusoidal and sawtooth profiles~\citep{Howe1991a,Howe1991}, see figure~\ref{fig:SerratedFlatPlate}. Howe's models show that sawtooth serrations are more effective in reducing the trailing-edge noise than the sinusoidal ones and that the use of sharp sawtooth serrations, i.e. $2h/\lambda > 8$, can lead to significant reduction of trailing-edge noise. More recently, \citet{Azarpeyvand2013} carried out an analytical investigation of trailing-edge noise reduction using novel serrations, namely, sawtooth, sinusoidal, slitted, slitted-sawtooth and sawtooth-sinusoidal. It has been found that the noise reduction is a sensitive function of the complexity of the serration geometry and significant noise reduction can be achieved by applying complex periodic serrations to the trailing-edge. It has also been shown that the slitted-sawtooth serration is the most effective design amongst the aforementioned serration geometries.

An experimental investigation on trailing-edge serrations was performed by \citet{Dassen1996}. Both aerofoils and flat plates of different shapes were tested in a wind tunnel. A maximum of 10 dB noise reduction for the flat plate and 8 dB reduction for the aerofoils were reported, both of which occurred mainly at low frequencies. Later, \citet{Parchen1999} conducted an experimental investigation of the aeroacoustic effects of trailing-edge serrations on wind turbine blades at both full and wind-tunnel scales. An average sound reduction slightly below that reported by Dassen {\it et al.} was observed. Most recently, \citet{Oerlemans2009} examined and compared the noise generated by standard, serrated and aeroacoustically-optimized aerofoils, but only 2-3 dB noise reduction was achieved for the aerofoil with serrated trailing-edge at low frequencies. Both \citet{Parchen1999} and \citet{Oerlemans2009} reported a noise increase at high frequencies.

\citet{Gruber2012} recently conducted an extensive experimental investigation on the aeroacoustic performance of aerofoils fitted with different sawtooth and novel serrations. The acoustic measurements were performed to give the sound power level (SWL) integrated in the mid-span plane. An average of 3-5 dB reduction was achieved using sharp sawtooth serrations, and a noise increase of up to 5 dB at higher frequencies was also reported. It was explained that the significant reduction of phase speed near the sawtooth edges, together with a slight reduction of the coherence of pressure measured along the edge is responsible for the sound reduction observed in experiments.  All the experimental studies, however, indicate that Howe's model significantly overpredicts the sound reduction capability of trailing-edge serrations. This might be caused by the assumptions and approximations used in Howe's derivation, which will be discussed in detail in the subsequent sections.

Although different serrations have been used in many applications such as wind turbines and jet nozzles~\citep{Oerlemans2009,Callender2005, Yan2007}, the physical mechanism of the noise reduction remains poorly understood. Howe's model gave a first insight into the physics involved, but the large deviation from experiments suggests a new and more accurate theory is needed.

The main objectives of this paper are to develop a new theory to predict the sound generated from a serrated trailing-edge more accurately, and to improve our understanding of the sound reduction mechanism due to the presence of trailing-edge serrations. The paper is structured as follows: the new theoretical model for sound radiation from serrated trailing-edges is presented in section~\ref{sec:II}. Section~\ref{sec:III} provides validation results against finite element solution. A parametric study will also be presented and the effects of serrations on trailing-edge noise directivity will also be discussed. A comparison between the new model and Howe's model is made in section~\ref{sec:IV}. The physical mechanism of sound reduction using trailing-edge serrations is discussed in section~\ref{sec:V} and noise reduction criteria developed based on the results in section~\ref{sec:II} will be listed and discussed. A brief conclusion is given in the last section.
\nomenclature[z-FEM]{FEM}{Finite Element Method}\nomenclature[a-L]{$\mathcal{L}$}{Gust-response function}
\nomenclature[z-SWL]{SWL}{Sound poWer Level}

\section{Analytical formulation}
\label{sec:II}
%In this section, we present a detailed derivation for the noise prediction of serrated trailing-edges.
As for Amiet's model, the analytical model developed here is based on Schwarzschild's technique for wave equation with a discontinuous boundary condition. It is therefore useful to begin with describing this technique. The Schwarzschild method~\citep{Amiet1976, Roger2005} states that if a function $f(x,y)$ satisfies
\begin{equation}
	\left\{
	\begin{aligned}
		& \frac{\partial^2 f}{\partial x^2} + \frac{\partial^2 f}{\partial y^2} + \mu^2 f = 0 & \\
		& \frac{\partial f}{\partial y}(x,0)= 0, &x<0 \\
		& f(x,0) = g(x), &x\ge0,
	\end{aligned}
	\right.
\end{equation}
then, for $x < 0$,
\begin{equation}
	f(x,0) = \frac{1}{\pi} \int_0^{\infty} \sqrt{\frac{-x}{\xi}} \frac{\text{e}^{\text{i}\mu(\xi-x)}}{\xi-x} g(\xi) \ud\xi.
	\label{equ:SchwarzschildSolution}
\end{equation}
As shown by \citet{Amiet1976b}, the above method can be used to obtain the scattered pressure field over the surface of the aerofoil.

\subsection{The mathematical model}
\label{sec:TheoreticalModelling}

Consider an aerofoil with trailing-edge serrations, modeled as a flat plate as shown in figure~\ref{fig:SerratedFlatPlate}, with an infinitesimal thickness and an averaged chord length $c$ and spanwise length $d$. Let $x^{\prime}$, $y^{\prime}$ and $z^{\prime}$ denote the streamwise, spanwise and normal to the plate coordinates, respectively. The observer point is located at $(x_1,x_2,x_3)$. The profile function $H(y^\prime)$ is used to describe the serrated edges. The origin of the coordinates is chosen in such a way that $H(y^\prime)$ is an oscillatory function of zero mean and that $H(y^\prime)= 0$ in the absence of serrations. Figure~\ref{fig:SerratedFlatPlate} shows  a sawtooth serration with a root-to-tip length of $2h$ and a periodic wavelength of $\lambda$.

\nomenclature[a-x]{$x^{\prime}$}{Chordwise axis in aerofoil-fixed frame}
\nomenclature[a-y]{$y^{\prime}$}{Spanwise axis in aerofoil-fixed frame}
\nomenclature[a-z]{$z^{\prime}$}{Axis perpendicular to aerofoil in aerofoil-fixed frame}
\nomenclature[a-c]{$c$}{Averaged chord length}
\nomenclature[a-d]{$d$}{Span length}
\nomenclature[a-x1]{$x_1$}{The projection of observer point on $x^{\prime}$}
\nomenclature[a-y1]{$x_2$}{The projection of observer point on $y^{\prime}$}
\nomenclature[a-z1]{$x_3$}{The projection of observer point on $z^{\prime}$}

\nomenclature[a-H]{$H(y^\prime)$}{Serration profile function}
\nomenclature[a-h]{$h$}{Half of root-to-tip length of serrations}
\nomenclature[g-lambda]{$\lambda$}{Wavelength of serrations}

When the sound wavelength is equal to or shorter than the chord length $c$, the flat plate can be treated as a semi-infinite plate without a leading-edge~\citep{Amiet1976}. Furthermore, the plate can be considered infinite in the spanwise direction provided it has a relatively large aspect ratio (typically $d/c > 3$)~\citep{Amiet1978,Roger2010} . The turbulence inside the boundary layer is assumed to be frozen, i.e. it remains statistically the same before and after passing over the trailing-edge.

After implementing a spatial and time Fourier transformation, the hypothetical surface pressure beneath the turbulent boundary layer that would exist when the flat plate is absent can be expressed as an integral of different wall pressure gust components. The incoming wall pressure gust of frequency $\omega$ convected at a speed $U_c$, as illustrated in figure~\ref{fig:SerratedFlatPlate}, takes the form of
\begin{equation}
	p_i = P_{i}\text{e}^{-\text{i}(\omega t-k_1 x^{\prime}-k_2 y^{\prime})},
	\label{equ:GustForSerrated}
\end{equation}
where $P_{i}$ is the magnitude of the incident wall pressure gust and $k_1$ and $k_2$ denote the wavenumbers in the chordwise and spanwise directions, respectively.

\nomenclature[a-i]{$\text{i}$}{Imaginary symbol $\sqrt{-1}$}
\nomenclature[a-p_i]{$p_i$}{Incident wall pressure gust}
\nomenclature[g-o]{$\omega$}{Angular frequency $\omega = 2\pi f$}
\nomenclature[a-k_1]{$k_1$}{Hydrodynamic wavenumber in chordwise direction}
\nomenclature[a-k_2]{$k_2$}{Hydrodynamic wavenumber in spanwise direction}
\nomenclature[a-t]{$t$}{Time}

\nomenclature[a-Uc]{$U_c$}{Turbulence convection velocity}
\nomenclature[a-Pia]{$P_{i}$}{Magnitude of incident wall pressure gust}

The sound sources due to the presence of solid boundaries~\citep{Curle1955} can be modelled as dipoles, in addition to the quadrupoles in free field~\citep{Lighthill1952}. As explained in Amiet's paper~\citep{Amiet1976}, the incident pressure produces a scattered field originating from the trailing-edge, due to the change in boundary condition at the wall. The scattered field induces a pressure jump that cancels the incident pressure jump at the trailing-edge and in the wake after the plate (Kutta condition). Thus, the total pressure can be decomposed into two parts, namely $p_t = p_i + p$. The incident wall pressure is given by \ref{equ:GustForSerrated} and the scattered pressure field, $p$, must satisfy the following conditions at $z^{\prime} = 0$,
\begin{equation}
	\left\{
	\begin{aligned}
		&\frac{\partial p}{\partial z^{\prime}} = 0, &x^{\prime}< H(y^{\prime})\\
		& p = -P_{i}\text{e}^{\text{-i}(\omega t - k_1x^{\prime} -k_2 y^{\prime})}, & x^{\prime} \ge H(y{^\prime}).
	\end{aligned}
	\label{equ:BoundaryConditionsForSerrated}
	\right.
\end{equation}
It is worth pointing out that the incident pressure $p_i$ and the scattered pressure $p$ on the surface of the plate are, as in Amiet's paper~\citep{Amiet1976}, identical to the incident pressure jump and the scattered pressure jump, respectively, as mentioned in the preceding paragraph. This is merely a mathematical simplification and can be shown to be valid through a more rigorous analysis.

In the plate-fixed frame \{$x^{\prime},y^{\prime},z^{\prime}$\}, the air flow has a uniform speed $U$ in the streamwise direction outside the boundary layer and the wave equation governing the scattered pressure field $p$ is
\begin{equation}
\nabla^2 p - \frac{1}{c_0^2} \left(\frac{\partial}{\partial t} + U \frac{\partial}{\partial x^{\prime}}\right)^2 p = 0,
	\label{equ:TimeWaveEquation}
\end{equation}
where $c_0$ denotes the speed of sound. With the assumption of harmonic perturbation $p = P(x^{\prime}, y^{\prime},z^{\prime})\text{e}^{\text{-i} \omega t}$, the above equation reduces to
\begin{equation}
    \beta^2 \frac{\partial^2 P}{\partial x^{\prime 2}} + \frac{\partial^2 P}{\partial y^{\prime 2}} +\frac{\partial^2 P}{\partial z^{\prime 2}} + 2ikM_0\frac{\partial P}{\partial x^{\prime}} + k^2 P = 0,
\end{equation}
where $k = \omega/c_0$, $\beta^2 = 1-M_0^2$ and $M_0 = U/c_0$.

\nomenclature[a-u]{$U$}{Uniform flow velocity}
\nomenclature[a-P]{$P$}{Time-Fourier transformation of scattered pressure}
\nomenclature[a-c]{$c_0$}{Speed of sound in air}
\nomenclature[a-p]{$p$}{Scattered pressure field}
\nomenclature[a-k]{$k$}{Acoustic wavenumber, $\omega/c_0$}
\nomenclature[g-beta]{$\beta$}{$\beta = \sqrt{1-M_0^2}$}
\nomenclature[a-M0]{$M_0$}{Mach number of uniform flow}
%\nomenclature[a-Pbar]{$\bar{P}$}{Factorized $P$ defined in (\ref{equ:Pbar})}

In order to make the boundary conditions in (\ref{equ:BoundaryConditionsForSerrated}) independent of $y^{\prime}$, the coordinate transformation~\citep{Roger2013} $x = x^{\prime} - H(y^{\prime}),\, y  = y^{\prime},\, z = z^{\prime}$ is used and leads to the following differential equation:
\begin{equation}
\left(\beta^2+{H^\prime}^2(y)\right)\frac{\partial^2 P}{\partial x^2} + \frac{\partial^2 P}{\partial y^2} + \frac{\partial^2 P}{\partial z^2} - 2H^{\prime}(y)\frac{\partial^2 P}{\partial x\partial y} + \left(2iM_0k-H^{\prime \prime}(y)\right)\frac{\partial P}{\partial x} + k^2P = 0,
\label{equ:StretchedWaveEquation}
\end{equation}
where $H^{\prime}(y)$ and $H^{\prime \prime}(y)$ denote the first and second derivatives of $H(y)$. The boundary conditions now read
\begin{equation}
	\begin{cases}
		P(x, y, 0) = - P_{i}\text{e}^{\text{i}(k_1 x + k_2 y)} \text{e}^{\text{i} k_1 H(y)}, & x \ge 0 \\
		\partial P(x, y, 0)/\partial z= 0, & x < 0. \\
\end{cases}
	\label{equ:StretchedBoundaryConditions}
\end{equation}

Since the coefficients in (\ref{equ:StretchedWaveEquation}) are $y$-dependent, the standard ``separation of variables'' technique cannot be applied to solve this equation. We therefore turn to using a Fourier expansion technique in the following derivation.

\nomenclature[a-x]{$x$}{Chordwise axis in stretched coordinate system}
\nomenclature[a-y]{$y$}{Spanwise axis in stretched coordinate system}
\nomenclature[a-z]{$z$}{Axis perpendicular to aerofoil in stretched coordinate system}

\subsection{Fourier expansion}
\label{sec:FloquetAndFourier}
%According to Floquet theorem~\citep{Floquet1883, Eastham1973}, the pressure field scattered by the periodic serrations can be expanded in the physical frame $(x^{\prime},y^{\prime},z^{\prime})$ as
%\begin{equation}
%	P = \sum_{m=-\infty}^{\infty} P_m(x^{\prime},z^{\prime}) \text{e}^{\text{i}(k_2 + 2m\pi/\lambda)y^{\prime}}.
%	\label{equ:Floquet}
%\end{equation}
%This means that the function $P\text{e}^{\text{-i}k_2y^{\prime}}$ is a periodic function with period $\lambda$ in the $y^{\prime}$ direction. Thus, in the stretched frame $(x,y,z)$, as $x=x^{\prime} - H(y^{\prime})$, $y = y^{\prime}$ and $z = z^{\prime}$, function $P\text{e}^{\text{-i}k_2y}$ is also a periodic function with period $\lambda$ in the $y$ direction.
As the scattering problem is periodic in the spanwise direction, one can expand the scattered pressure field using Fourier series as
\begin{equation}
	P(x,y,z) = \sum_{-\infty}^{\infty}P_n(x,z)\text{e}^{\text{i}k_{2n}y},
	\label{expansion}
\end{equation}
where $k_{2n} = k_2 + 2n\pi/\lambda$.

\nomenclature[a-Pn]{$P_n$}{Scattered pressure of mode $n$}
\nomenclature[a-k2n]{$k_{2n}$}{Characteristic wavenumber of mode $n$, $k_{2n} = k_2 + 2 n \pi/\lambda$}
\nomenclature[r-']{$\square^\prime$}{First derivative of function}
\nomenclature[r-'']{$\square^{\prime\prime}$}{Second derivative of function}

Substituting the above expression into the transformed wave equation, shown in (\ref{equ:StretchedWaveEquation}), yields
\begin{equation}
	\begin{aligned}
		\left\{\left(\beta^2+{H^\prime}^2(y)\right)\frac{\partial^2 }{\partial x^2} + \frac{\partial^2 }{\partial y^2} + \frac{\partial^2 }{\partial z^2} - 2H^{\prime}(y)\frac{\partial^2 }{\partial x\partial y} + \left(2\text{i}M_0k-H^{\prime \prime}(y)\right)\frac{\partial }{\partial x} + k^2\right\}\\
	\sum_{-\infty}^{\infty}P_n(x,z)\text{e}^{\text{i}k_{2n}y} = 0.
\end{aligned}
\label{Ex}
\end{equation}
Multiplying (\ref{Ex}) by $\text{e}^{\text{-i}k_{2n^{\prime}}y}$, then integrating it over $y$ from $-\lambda/2$ to $\lambda/2$, it can be readily shown that
\begin{equation}
	\begin{aligned}
		&\left\{	\beta^2\frac{\partial^2}{\partial x^2} \right. + \left.\frac{\partial^2 }{\partial z^2} + 2\text{i}kM_0\frac{\partial }{\partial x} +  (k^2 - k_{2n^{\prime}}^2) \right\}P_{n^{\prime}}(x, z) + \\
                &  \frac{1}{\lambda}\int_{-\lambda/2}^{\lambda/2} \sum_{n = -\infty}^{\infty} \left\{  H^{\prime 2}(y) \frac{\partial^2}{\partial x^2} -(H^{\prime \prime}(y)+2\text{i}k_{2n} H^{\prime}(y))\frac{\partial}{\partial x} \right\}P_n(x,z) \text{e}^{\text{i}[2(n-n^{\prime})\pi/\lambda] y} \ud y=0.
		\label{equ:GeneralCoupledEquation}
	\end{aligned}
\end{equation}

Note that when both $H^{\prime}(y)$ and $H^{\prime \prime}(y)$ are constant within an entire sawtooth wavelength, the summation over different modes in (\ref{equ:GeneralCoupledEquation}) can be dropped and we obtain a fully decoupled differential equation for mode $n^{\prime}$. However, this means that the flat plate has a straight or swept trailing-edge. For serrations of an arbitrary profile, both $H^{\prime}(y)$ and $H^{\prime\prime}(y)$ generally depend on $y$. Thus~\ref{equ:GeneralCoupledEquation} becomes a coupled differential equation, i.e. more than one mode appears in each differential equation. The physical interpretation of this mode coupling will be discussed later.

\begin{figure}
    \centering
    \includegraphics[width=.5\textwidth]{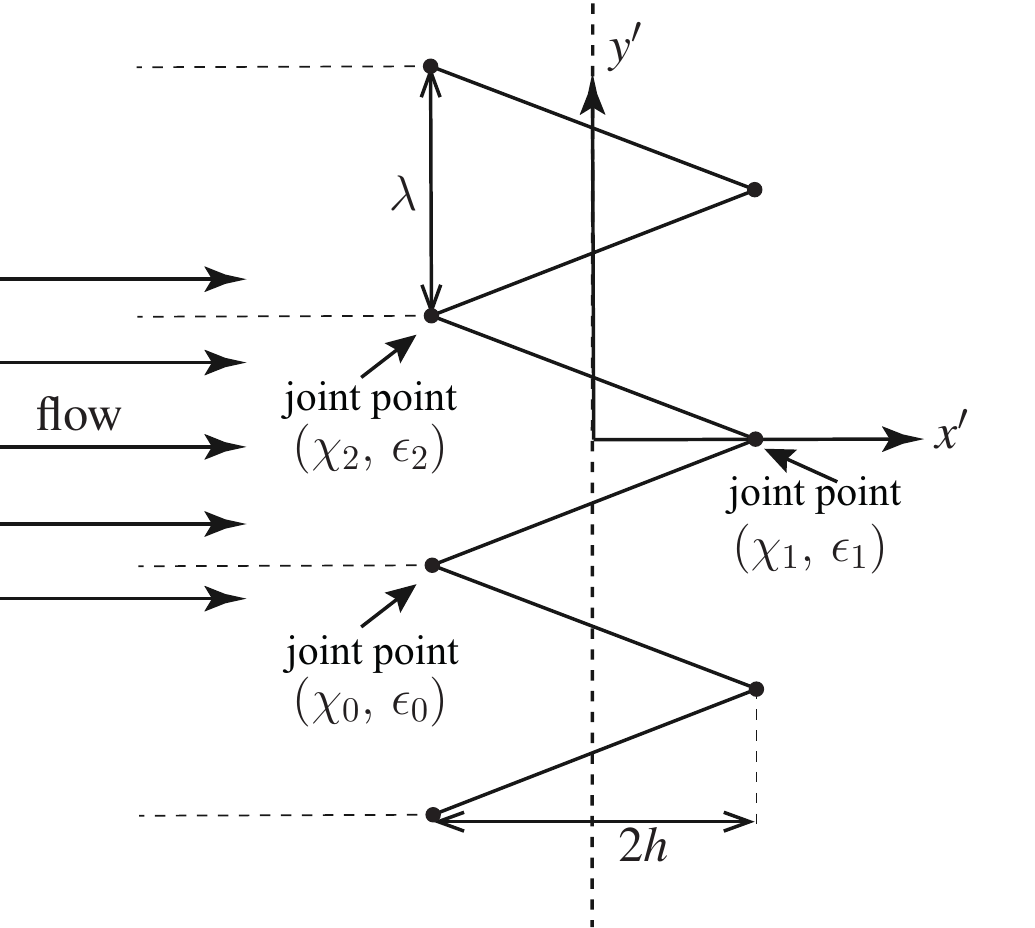}
    \caption{The schematic of sawtooth serrations}
    \label{fig:serrationProfile}
\end{figure}

In this paper, we only focus on the sawtooth serration, which has been shown to be effective in reducing the trailing-edge noise~\cite{Howe1991}. The method can however also be used for other serrations. Consider a sawtooth centred around the coordinate origin, and let $(\chi_0, \epsilon_0)$, $(\chi_1, \epsilon_1)$ and $(\chi_2, \epsilon_2)$ be the Cartesian coordinates of the tip and roots of the sawtooth, as shown in figure~\ref{fig:serrationProfile}.
The serration profile function $H(y)$ can therefore be defined as
\begin{equation}
	H(y) =
	\begin{cases}
		\sigma_0(y - \chi_0 - m\lambda) + \epsilon_0,  & \chi_0+m\lambda < y \le \chi_{1}+m\lambda\\
		\sigma_1(y - \chi_1 - m\lambda) + \epsilon_1,  & \chi_1+m\lambda < y \le \chi_{2}+m\lambda, \\
	\end{cases}
	\label{equ:ProfileFunction}
\end{equation}
where $\sigma_j = (\epsilon_{j+1} - \epsilon_j)/(\chi_{j+1} - \chi_j)$, $j = 0, 1$ and $m = 0, \pm1, \pm2, \pm3\cdots$.
Let $\sigma$, defined as $\sigma = |\sigma_j| = 4h/\lambda$, denote the sharpness of the sawtooth serrations. For a sawtooth profile, as mentioned above, $H^{\prime}(y)$ is not continuous and $H^{\prime \prime}(y)$ is thus singular at the joint-points. We use the conventional generalized function $\delta(x)$ to describe the singularities at these points, i.e.
\begin{equation}
	\begin{aligned}
		&	H^{\prime}(y) =
		\begin{cases}
		\sigma_0 ,  & \chi_0+m\lambda < y \le \chi_{1}+m\lambda\\
		\sigma_1,  & \chi_1+m\lambda < y \le \chi_{2}+m\lambda\\
		\end{cases}\\
		&H^{\prime \prime}(y) = \sum_{m = -\infty}^{\infty} (-1)^{m+1}2\sigma \delta(x-m\lambda/2).
	\end{aligned}
	\label{equ:1stDerivative}
\end{equation}

\nomenclature[g-simga]{$\sigma$}{Sawtooth serration sharpness factor, $4h/\lambda$}
\nomenclature[s-j]{$\square_j$}{The $j$-th edge of a single piece of sawtooth serrations}
\nomenclature[g-lambdaj]{$\chi_j$}{The spanwise coordinate of the starting point of the $j$-th edge }
\nomenclature[g-epsilonj]{$\epsilon_j$}{The chordwise coordinate of the starting point of the $j$-th edge}
\nomenclature[g-lambdajp]{$\chi_{j+1}$}{The spanwise coordinate of the ending point of the $j$-th edge }
\nomenclature[g-epsilonjp]{$\epsilon_{j+1}$}{The chordwise coordinate of the ending point of the $j$-th edge}
\nomenclature[g-sigmaj]{$\sigma_j$}{Slope of the $j$-th edge }

Substituting the serration profile function and its derivatives, (\ref{equ:ProfileFunction}) and (\ref{equ:1stDerivative}), into the wave equation, (\ref{equ:GeneralCoupledEquation}), and making use of the fact that $\int_{-\infty}^{\infty} f(x) \delta(x-\tau) \ud x = f(\tau)$, we obtain
\begin{equation}
	\begin{aligned}
		\left\{\left(\beta^2+\sigma^2\right) \frac{\partial^2}{\partial x^2} + \frac{\partial^2}{\partial z^2} + \right. & \left. 2\text{i}kM_0 \frac{\partial}{\partial x} + (k^2-k_{2n^{\prime}}^2)\right\} P_{n^{\prime}}(x, z) \\
	= & - \frac{4\sigma}{\lambda}\sum_{n-n^{\prime} = odd}\left(1-\frac{k_2\lambda + 2n\pi}{(n-n^{\prime})\pi}\right)\frac{\partial P_{n}(x, z)}{\partial x}.
	\label{coupled}
	\end{aligned}
\end{equation}
To make the above equation more compact, we write the set of differential equations obtained above into a matrix form. Let a linear operator
\begin{equation}
	\mathrm{D} = \left\{\left(\beta^2+\sigma^2\right) \frac{\partial^2}{\partial x^2} + \frac{\partial^2}{\partial z^2} + 2\text{i}kM_0 \frac{\partial}{\partial x}  \right\},
	\label{equ:Operator}
\end{equation}
and a vector of functions
\begin{equation}
	\boldsymbol{P} = \left(
	\cdots P_{-n^{\prime}}(x,z)\text{, }
	P_{-n^{\prime}+1}(x,z)\text{, }\cdots
		P_{n^{\prime}-1}(x,z)\text{, }
		P_{n^{\prime}}(x,z)\text{, }\cdots
	\right)^T,
	\label{equ:VectorP}
\end{equation}
then the coupled equations can be written as
\begin{equation}
	\mathrm{D}\boldsymbol{P} - \mathsfbi{A} \boldsymbol{P} = \mathsfbi{B}\frac{\partial \boldsymbol{P}}{\partial x},
	\label{equ:MatrixFormWaveEquations}
\end{equation}
where the symbol $T$ in (\ref{equ:VectorP}) represents the transpose of a matrix. Matrices $\mathsfbi{A}$ and $\mathsfbi{B}$ denote the coefficient matrices of $\boldsymbol{P}$ and $\partial \boldsymbol{P}/\partial x$, respectively. $A_{ml}$ and $B_{ml}$ representing the entry corresponding to mode $m$ in row and $l$ in column of matrices $\mathsfbi{A}$ and $\mathsfbi{B}$ are
\begin{equation}
	\begin{aligned}
		&A_{ml} = (k_{2m}^2 - k^2)\delta_{ml},
		&B_{ml} = \Bigg\{
			\begin{aligned}
                            &\frac{4\sigma}{\lambda}\frac{m+l+k_2\lambda/\pi}{l-m}, &m-l \text{ is odd}\\
                            &0, &m-l \text{ is even},
			\end{aligned}
		\end{aligned}
		\label{equ:ABDefinition}
	\end{equation}
        where $\delta_{ml}$ represents the Kronecker delta. It is worth noting that the index of matrices $\mathsfbi{A}$ and $\mathsfbi{B}$ are from $-m$ to $m$ and $-l$ to $l$ rather than from $1$ to $2m+1$ and $1$ to $2l+1$ given the fact that the mode number are symmetric with respect to $0$.

\nomenclature[a-D]{$\mathrm{D}$}{Linear differential operator}
\nomenclature[a-P]{$\boldsymbol{P}$}{Vector expression of scattered pressure of different modes}
\nomenclature[a-A]{$\mathsfbi{A}$}{Coefficient matrix for $\boldsymbol{P}$ defined in (\ref{equ:MatrixFormWaveEquations})}
\nomenclature[a-B]{$\mathsfbi{B}$}{Coefficient matrix for $\partial \boldsymbol{P}/\partial x$ defined in (\ref{equ:MatrixFormWaveEquations})}
\nomenclature[a-A]{$A_{ml}$}{Elements of matrix $\mathsfbi{A}$}
\nomenclature[a-B]{$B_{ml}$}{Elements of matrix $\mathsfbi{B}$}
\nomenclature[g-detla]{$\delta_{ml}$}{The Kronecker delta}
\nomenclature[g-tau]{$\tau$}{Variable in function $\delta(x-\tau)$}

Substituting the profile geometry, (\ref{equ:ProfileFunction}), into the boundary conditions, (\ref{equ:StretchedBoundaryConditions}), and performing the same Fourier expansions, yields
\begin{equation}
	\left\{
		\begin{aligned}
		        &P_n(x,0) = -P_{i}a_n\text{e}^{\text{i}k_1x} ,& x>0\\
			&\frac{\partial P_n}{\partial z}(x,0) = 0, &x \le 0,
		\end{aligned}
		\label{equ:MatrixFormBoundaryConditions}
		\right.
\end{equation}
where $a_n$ are defined as
\begin{equation}
	a_n = \frac{1}{\lambda}\int_{-\lambda/2}^{\lambda/2} \text{e}^{\text{i}k_1 H(y)} \text{e}^{\text{-i}2n\pi/\lambda y} \ud y.
	\label{an}
\end{equation}

\nomenclature[a-an]{$a_n$}{Fourier coefficient of mode $n$}

In (\ref{equ:MatrixFormWaveEquations}), as $\mathsfbi{B}$ is not a diagonal matrix, the term $\mathsfbi{B}(\partial \boldsymbol{P}/\partial x)$ contains coupling terms, in the sense that different modes, $P_{n}$ for example, appear in the governing equation of $P_{m}$. This means that every mode is interacting with the other modes and cannot be solved independently. From the expression of $\mathsfbi{B}$ in (\ref{equ:ABDefinition}), it can be seen that the strength of coupling is propotional to $\sigma/\lambda$. This indicates that shaper serrations have stronger coupling between different modes. Also note that $\mathsfbi{A}$ is a diagonal matrix, so if $\mathsfbi{B} \approx 0$, i.e. the serrations are very wide, then there is no coupling effect and one can solve each mode independently.
%It is worth pointing out that if the coupling terms are left out completely, then at certain frequencies where the modes contributing the most to the sound radiation disappear, a relatively sharp peak in sound reduction curve would appear. And the coupling effect tends to chop all the sharp peaks and to make sound reduction curve smoother.

At very low frequencies, the contribution of higher modes is expected to become gradually negligible compared to mode $0$. Thus, it can be reasonably assumed that in the case of the governing equation of $P_0$ (mode 0), the coupling with higher modes is weak and one can solve $P_0$ individually. The coupling effect becomes more pronounced at high frequencies and for sharp serrations. To solve these coupled equations at relatively high frequencies, one can use an iterative procedure, to be introduced in the next section.

%first ignore the coupling effects completely, and obtain an initial set of solutions $\boldsymbol{P}^{(0)}$. Then, substituting these initial values into the coupling terms in (\ref{equ:MatrixFormWaveEquations}) enables to solve a new set of solutions, $\boldsymbol{P}^{(1)}$. Continuing these steps gives a solution sequence, and if this sequence appears to converge to a set of functions, $\boldsymbol{P}$ for example, then $\boldsymbol{P}$ must be the solutions to the coupled differential equations shown in (\ref{equ:MatrixFormWaveEquations}).

\nomenclature[a-Pi]{$\boldsymbol{P}^{(i)}$}{The $i$-th iterated solutions of PDEs}

\subsection{The iterative solution procedure}
% In this section, we attempt to solve the coupled equation obtained in the preceding section and investigate analytically the effects of trailing-edge serrations.
\subsubsection{Scattered surface pressure}
\label{sec:ExtensionOfAmiet}
%\paragraph{\it 1. Iteration procedure:}
To obtain the scattered surface pressure, equation (\ref{equ:MatrixFormWaveEquations}) together with the boundary conditions in (\ref{equ:MatrixFormBoundaryConditions}) need to be solved. By analogy with the solution of a system of linear algebraic equation~\citep{Suli2003}, which can be obtained iteratively, we shall solve our system of PDEs in an iterative manner.

Substituting a known initial value $\boldsymbol{P}^{(0)}$ into the coupling term in (\ref{equ:MatrixFormWaveEquations}), one can write
\begin{equation}
	\mathrm{D}\boldsymbol{P} - \mathsfbi{A} \boldsymbol{P} = \mathsfbi{B}\frac{\partial \boldsymbol{P}^{(0)}}{\partial x}.
	\label{equ:iteration1st}
\end{equation}
Solving (\ref{equ:iteration1st}) yields a new set of solutions $\boldsymbol{P}^{(1)}$. Replacing $\boldsymbol{P}^{(0)}$ in (\ref{equ:iteration1st}) with $\boldsymbol{P}^{(1)}$, we obtain a new wave equation,
\begin{equation}
	\mathrm{D}\boldsymbol{P} - \mathsfbi{A} \boldsymbol{P} = \mathsfbi{B}\frac{\partial \boldsymbol{P}^{(1)}}{\partial x}.
	\label{equ:iteration2nd}
\end{equation}
Solving (\ref{equ:iteration2nd}) gives a new set of solutions $\boldsymbol{P}^{(2)}$. This process is repeated to obtain a solution sequence, $\boldsymbol{P}^{(0)}$, $\boldsymbol{P}^{(1)}$, $\boldsymbol{P}^{(2)}$, $\boldsymbol{P}^{(3)}\cdots$. If the sequence is convergent, then its limit satisfies (\ref{equ:MatrixFormWaveEquations}).

The initial value $\boldsymbol{P}^{(0)}$ used to start the first iteration can be obtained, as mentioned in the previous section, by ignoring all the coupling terms, i.e. with $\mathsfbi{B} = 0$, and by solving each equation individually via the standard Schwarzschild technique, as shown in (\ref{equ:SchwarzschildSolution}). The solution to each equation in the decoupled system of equations
\begin{equation}
	\mathrm{D}\boldsymbol{P} - \mathsfbi{A} \boldsymbol{P} = 0
	\label{initial}
\end{equation}
can be found as follows.

After use is made of the transformation of $P_{n^\prime} = \bar{P}_{n^\prime} \text{e}^{-\text{i}kM_0/(\beta^2 + \sigma^2)x}$, individual equations in (\ref{initial}) reduce to,
\begin{equation}
    \{(\beta^2 + \sigma^2) \frac{\partial^2}{\partial x^2} + \frac{\partial^2}{\partial z^2} + K_{n^\prime}^2(\beta^2 + \sigma^2)\} \bar{P}_{n^\prime} = 0,
    \label{equ:initialEle}
\end{equation}
where
\begin{equation}
		K_{n^{\prime}} = \sqrt{k^2(1+\sigma^2) - k_{2n^{\prime}}^2(\beta^2+\sigma^2)}/(\beta^2+\sigma^2).
    \label{equ:Kndefinition}
\end{equation}
Making use of $X = x, Z = \sqrt{\beta^2 + \sigma^2}z$, one can show that (\ref{equ:initialEle}) reduces to a standard Schwarzschild problem and the solution can be found using the Schwarzschild integral described in (\ref{equ:SchwarzschildSolution}), as
\begin{equation}
  P^{(0)}_{n^{\prime}} = P_{i} \text{e}^{\text{i}k_1 x} a_{n^{\prime}} \big((1-\text{i})\mathrm{E}(-\mu_{n^{\prime}} x) -1\big),
        \label{equ:initialResults}
\end{equation}
where $P^{(0)}_{n^{\prime}}$ is the element of the vector $\boldsymbol{P}^{(0)}$ corresponding to the $n^{\prime}$-th mode, and
\begin{equation}
    \begin{aligned}
            &\mu_{n^{\prime}} = K_{n^{\prime}} + k_1 +\frac{kM_0}{\beta^2 + \sigma^2},\\
            &\mathrm{E}(x) = \int_0^x \frac{\text{e}^{\text{i}t}}{\sqrt{2\pi t}} \ud t.
        \end{aligned}
\end{equation}
The initial solutions obtained by ignoring all the coupling terms denote the decoupled part of the exact solution of each mode, which implies that the $n$-th mode excitation ($x > 0$) produces only an $n$-th mode response ($x < 0$). The iteration procedure will add a coupled part to the solution of each mode. The coupled part implies that an $n$-th mode input ($x>0$) will also produce some $m$-th mode responses ($x<0$), where $m \ne n$. It can be expected that those coupling contributions from closer modes are stronger than remote ones. For sawtooth serrations, as will be shown in (\ref{equ:vdefinition}), the coupling strength decays quadratically with respect to the difference between their ``mode number'', i.e. $|m-l|$.

\nomenclature[a-E]{$\mathrm{E}(x)$}{Complex error function}
\nomenclature[z-PDE]{PDE}{Partial Differential Equation}
\nomenclature[a-P0]{$P_{n^{\prime}}^{(i)}$}{Elements of vector $\boldsymbol{P}^{(i)}$}
\nomenclature[g-mun]{$\mu_{n^{\prime}}$}{Transformed mixed wavenumber of mode $n^{\prime}$}
\nomenclature[a-Kn]{$K_{n^{\prime}}$}{Transformed acoustic wavenumber of mode $n^{\prime}$}
\nomenclature[a-P]{$\bar{P}_{n^\prime}$}{Transformed scattered modal pressure}

By substituting $\boldsymbol{P}^{(0)}$ into the coupling terms on the right hand side of (\ref{equ:MatrixFormWaveEquations}), one obtains some inhomogeneous equations that can no longer be solved using the standard Schwarzschild technique. However, if one can transform these equations into homogeneous ones, then Schwarzschild's method can again be applied. Note that $\boldsymbol{P}^{(0)}$ satisfies (\ref{initial}), hence, for $x \ne 0$, where $\boldsymbol{P}^{(0)}$ is first-order continuously differentiable, the following equation holds:
\begin{equation}
	\mathrm{D}\frac{\partial\boldsymbol{P}^{(0)}}{\partial x} - \mathsfbi{A} \frac{\partial\boldsymbol{P}^{(0)}}{\partial x} = 0.
	\label{initial_derivative}
\end{equation}
Making use of (\ref{initial_derivative}), (\ref{equ:iteration1st}) can be equivalently written as
\begin{equation}
	\mathrm{D} \left(\boldsymbol{P} + \mathsfbi{v}\frac{\partial \boldsymbol{P}^{(0)}}{\partial x}\right) - \mathsfbi{A}\left(\boldsymbol{P} + \mathsfbi{v} \frac{\partial \boldsymbol{P}^{(0)}}{\partial x}\right) = 0,
	\label{equ:Homogenous1st}
\end{equation}
where $\mathsfbi{v}$ is a coefficient matrix whose entries are
\begin{equation}
v_{ml} = \frac{B_{ml}}{k_{2m}^2 - k_{2l}^2} =
\Bigg\{
    \begin{aligned}
        &\frac{-4h}{\pi^2(m-l)^2}, &m-l = odd\\
        &0, &m-l = even.
    \end{aligned}
    \label{equ:vdefinition}
\end{equation}

It is worth pointing out that (\ref{equ:Homogenous1st}) only holds when $x \in \mathbf{R}$ and $x \ne 0$, and in order to apply the Schwarzschild technique, it must be valid over the whole domain. However, since the singularity of $\partial \boldsymbol{P}^{(0)}/\partial x$ only exists at $x = 0$, similar to the differentiation of $H(y)$, we may again make use of the generalized function to account for this singularity. Let $\partial \boldsymbol{\hat{P}}^{(0)}/\partial x$ denote the generalized differentiation, which allows the presence of generalized functions at singular point $ x= 0$ but equals to $\partial \boldsymbol{P}^{(0)}/\partial x$ elsewhere, then equation
\begin{equation}
	\mathrm{D}\frac{\partial\boldsymbol{\hat{P}}^{(0)}}{\partial x} - \mathsfbi{A} \frac{\partial\boldsymbol{\hat{P}}^{(0)}}{\partial x} = 0
	\label{equ:ModifiedP0}
\end{equation}
needs to hold over $x \in \mathbf{R}$. The Schwarzschild technique suggests that if (\ref{equ:ModifiedP0}) does hold, then the routine application of the steps described from (\ref{equ:initialEle}) to (\ref{equ:initialResults}) shall recover the value of $\partial \boldsymbol{P}^{(0)}/\partial x$ for $x < 0$. Thus, one can verify that the intended $\partial \boldsymbol{\hat{P}}^{(0)}/\partial x$ can indeed be found as
\begin{equation}
  \frac{\partial \hat{P}_{n^{\prime}}^{(0)}}{\partial x} (x,0)= \frac{\partial P_{n^{\prime}}^{(0)}}{\partial x}(x,0) + P_{i} a_{n^\prime}(1-\text{i})(-\sqrt{\mu_{n^{\prime}}}) \sqrt{2\pi x} \delta(x),
\label{Fexpression}
\end{equation}
where $\partial \hat{P}_{n^{\prime}}^{(0)}/\partial x$ denotes the element of $\partial \boldsymbol{\hat{P}}^{(0)}/\partial x$, corresponding to the $n^{\prime}$-th mode and
\begin{equation}
    \int_0^{\infty} \delta(x) \ud x = \frac{1}{2}.
	\label{equ:DeltaDefinition}
\end{equation}
%The $1/2$ factor in (\ref{equ:DeltaDefinition}) is defined to account for the integral over the interval from $0$ to $\infty$, which is only a half of the conventional interval from $-\infty$ to $\infty$.
Now, the first iterated solution can be obtained by solving equation
\begin{equation}
	\mathrm{D} \left(\boldsymbol{P} + \mathsfbi{v}\frac{\partial \boldsymbol{P}^{(0)}}{\partial x}\right) - \mathsfbi{A}\left(\boldsymbol{P} + \mathsfbi{v} \frac{\partial \boldsymbol{P}^{(0)}}{\partial x}\right) = 0,
	\label{equ:IterationEqu}
\end{equation}
via the steps described from (\ref{equ:initialEle}) to (\ref{equ:initialResults}).
%\begin{equation}
%    P^{(1)}_{n^{\prime}}(x,0) = \frac{1}{\pi} \int_0^{\infty} \sqrt{\frac{-x}{\xi}} \frac{\text{e}^{\text{i}\mu_{n^{\prime}}(\xi-x)}}{\xi-x}\left(P_{n^{\prime}} + \sum_{m^\prime} \alpha_{n^{\prime}m^\prime}\frac{\partial \hat{P}^{(0)}_{m^\prime}}{\partial \xi}\right)(\xi,0) \ud\xi - \sum_{m^\prime}\alpha_{n^{\prime}m^\prime}\frac{\partial \hat{P}^{(0)}_{m^\prime}}{\partial x}.
%	\label{firstiteration}
%\end{equation}

\nomenclature[a-v]{$\mathsfbi{v}$}{Coefficient matrix denoting coupling strength defined in (\ref{equ:vdefinition})}
\nomenclature[g-xi]{$\xi$}{Integral variable}
\nomenclature[a-Phat]{$\boldsymbol{\hat{P}}^{(0)}$}{Modified pressure in vector form}

Solving (\ref{equ:IterationEqu}) gives the values of $\boldsymbol{P}^{(1)}$. Continuing this iteration process gives $\boldsymbol{P}^{(2)}$, $\boldsymbol{P}^{(3)} \cdots$. The exact solutions $\boldsymbol{P}$ can also be expressed as
\begin{equation}
	\boldsymbol{P}(x,0) = \boldsymbol{N}(x) + \boldsymbol{C}^{(1)}(x)+ \boldsymbol{C}^{(2)}(x)+ \boldsymbol{C}^{(3)}(x)+ \cdots,
	\label{equ:Solutions}
\end{equation}
where $\boldsymbol{N}$ is the non-coupled part and the coupled parts are denoted by $\boldsymbol{C}^{(i)} = \boldsymbol{P}^{(i)} - \boldsymbol{P}^{(i-1)}$ ($i = 1, 2, 3\cdots$). The entries of $\boldsymbol{N}$ and $\boldsymbol{C}^{(1)}$ corresponding to mode $n^{\prime}$ are given by
\begin{equation}
    N_{n^{\prime}}(x) = P_{i} \text{e}^{\text{i}k_1x} a_{n^{\prime}} \left( (1-\text{i})\mathrm{E}(-\mu_{n^{\prime}}x) - 1 \right),
	\label{equ:NonscatteringPart}
\end{equation}
\begin{equation}
	\begin{aligned}
            C_{n^{\prime}}^{(1)}(x) = P_{i}\text{e}^{\text{i}k_1x}(1-\text{i}) \sum_{m=-\infty}^\infty v_{n^{\prime}m}a_m \Bigg(\text{i}k_1 \Big( \mathrm{E}(-\mu_{n^{\prime}}x)-\mathrm{E}(-\mu_m x)\Big)&\\
	  -\sqrt{\frac{\mu_m}{-2\pi x}}\Big(\text{e}^{\text{-i}\mu_{n^{\prime}}x} - \text{e}^{\text{-i}\mu_m x} \Big)\Bigg).&
	\end{aligned}
	\label{equ:ScatteringPart1}
\end{equation}
The elements of the second-order function $\boldsymbol{C}^{(2)}$ are provided in appendix A. Since $v_{ml} \propto h$ it can be readily shown that $C^{(i)}_{n^\prime} \propto h^i$ for $i = 1,\,2,\,3\cdots$. This means that the solution presented in (\ref{equ:Solutions}) is a perturbation (Taylor-expansion) series with respect to half of the root-to-tip length $h$. Therefore a smaller value of $h$ compared to the sound wavelength yields faster convergence. It is worth noting that the function $\boldsymbol{C}^{(i)}$ becomes more and more complex as $i$ increases. However, if $\boldsymbol{C}^{(i)}$ vanishes sufficiently quickly, higher orders can be neglected without causing significant errors. This appears to be the case for the frequencies relevant to trailing-edge noise, see Sec.~\ref{subsec:Results}. Substituting (\ref{equ:NonscatteringPart}), (\ref{equ:ScatteringPart1}) and (\ref{equ:secondOrderPart})  into (\ref{equ:Solutions}), a second-order approximation of the exact solutions is obtained.
\nomenclature[a-N]{$\boldsymbol{N}$}{Non-coupled part of solution in $\boldsymbol{P}$}
\nomenclature[a-C]{$\boldsymbol{C}^{(i)}$}{Coupled part of solution in $\boldsymbol{P}$}
\nomenclature[a-N]{$N_{n^{\prime}}$}{Elements of vector $\boldsymbol{N}$}
\nomenclature[a-C]{$C_{n^{\prime}}^{(i)}$}{Elements of vector $\boldsymbol{C}^{(i)}$}

%\paragraph{\it 3. Scattered surface pressure }
The scattered surface pressure is obtained by summing $P_{n^\prime}(x, 0)$ over all different modes and transforming  back to the physical coordinate system, namely
\begin{equation}
	P(x^{\prime},y^{\prime},0)= \sum_{n^{\prime}=-\infty}^{\infty} P_{n^{\prime}}(x^{\prime}-H(y^{\prime}),0) \text{e}^{\text{i}k_{2n^{\prime}}y^{\prime}}.
	\label{equ:ScatteredSurfacePressure}
\end{equation}
Here, $P_{n^{\prime}}$ is the solution obtained from the iteration procedure mentioned above,
\begin{equation}
	\begin{aligned}
		P(x^{\prime},y^{\prime},0) =\sum_{n^{\prime}=-\infty}^{\infty} (N_{n^{\prime}}+C^{(1)}_{n^{\prime}} + C^{(2)}_{n^{\prime}}+ \cdots)(x^{\prime}-H(y^{\prime}),0) \text{e}^{\text{i}k_{2n^{\prime}}y^{\prime}},
		\label{equ:ScatteredSurfacePressure1st}
	\end{aligned}
\end{equation}
where $N_{n^{\prime}}$ and $C^{(1)}_{n^{\prime}}$ are defined in (\ref{equ:NonscatteringPart}) and (\ref{equ:ScatteringPart1}), and $C^{(2)}_{n^{\prime}}$ can be found in appendix A. Note that the terms in the second parenthesis are the arguments for the $N_{n^\prime}$ and $C^{(i)}_{n^\prime} \,(i = 1, 2, 3, \cdots)$ functions.  It is worth pointing out that in the limiting case when $H(y^\prime) =0$, $\boldsymbol{C}^{(i)}$ vanishes, and (\ref{equ:ScatteredSurfacePressure1st}) reduces to the result obtained by \citet{Amiet1976} for a straight edge.

%\paragraph{\it 4. Convergence of infinite series}
As shown in (\ref{equ:ScatteredSurfacePressure1st}), the scattered pressure field can now be expressed in terms of an infinite series. By inspection of (\ref{equ:ScatteredSurfacePressure1st}), one can show that at sufficiently low frequencies, i.e. $k_1 h < \pi^2 /4$, the infinite series is absolutely convergent. At higher frequencies, the series still appears to be convergent, but to obtain satisfactory approximation, a higher truncation number and higher-order iterations may be required. The convergence of the series will be disscussed in the following sections by comparing the far-field sound prediced using different order approximations.

\subsubsection{Far-field sound pressure}
\label{subsubsec:far-fieldPressure}
As illustrated in figure~\ref{fig:SerratedFlatPlate}, the observer point is located at $(x_1,x_2,x_3)$ and the flat plate has an averaged chord length $c$ and span length $d$. The far field sound can be found using the surface pressure integral as mentioned in Amiet's model~\citep{Lamb1932,Curle1955,Amiet1975}:
\begin{equation}
	p_f(\mathbf{x}, \omega) = \frac{-\text{i}\omega x_3}{4\pi c_0 S_0^2} \iint_s \Delta P(x^{\prime}, y^{\prime}) \text{e}^{\text{-i}kR} \ud x^{\prime} \ud y^{\prime},
	\label{equ:SerratedFarField}
\end{equation}
where $\Delta P = P$ denotes the pressure jump, and $S_0^2 = x_1^2 + \beta^2(x_2^2 + x_3^2)$ and
\begin{equation}
	R= \frac{M_0(x_1 - x^{\prime})-S_0}{\beta^2} + \frac{x_1x^{\prime} + x_2 y^{\prime}\beta^2}{\beta^2S_0}.
	\label{equ:PhaseRelation}
\end{equation}
%is taken to account for the opposite pressure distribution on upper and lower sides of the aerofoil.

\nomenclature[a-S]{$S_0$}{$S_0 = \sqrt{x_1^2 + \beta^2(x_2^2 + x_3^2)}$}
\nomenclature[a-R]{$R$}{Function denoting phase relation shown in (\ref{equ:PhaseRelation})}
\nomenclature[a-pdelta]{$\Delta P$}{Pressure jump across the aerofoil}

By substituting the solution obtained in (\ref{equ:ScatteredSurfacePressure1st}) into (\ref{equ:SerratedFarField}), the far-field sound pressure can be expressed as
\begin{equation}
	p_f(\mathbf{x},\omega,k_2) = P_{i}\left(\frac{-\text{i}\omega x_3c}{4\pi c_0 S_0^2}\right)\lambda\frac{\sin\big((N+1/2)\lambda(k_2-kx_2/S_0)\big)}{\sin\big(\lambda/2(k_2-kx_2/S_0)\big)}\mathcal{L}(\omega,k_2).
	\label{equ:PSD}
\end{equation}
Here, $2N+1$ represents the number of sawteeth on the edge and the far-field sound gust-response function $\mathcal{L}$ is defined as
\begin{equation}
	\mathcal{L}(\omega,k_2) = (1-\text{i})\frac{1}{\lambda c}\left(\sum_{n^{\prime }=-\infty}^\infty \big(\Theta_{n^{\prime}} + \Theta_{n^{\prime}}^{(1)} +\Theta_{n^{\prime}}^{(2)}+\cdots \big)\right)\text{e}^{\text{-i}k/\beta^2(M_0x_1-S_0)} \text{e}^{\text{i}k/\beta^2(M_0-x_1/S_0)h},
	\label{equ:SerratedResponseFunction}
\end{equation}
with (only the first two terms are given, see more results in appendix A)
\begin{equation}
\label{eq:Thetan'}
	\begin{aligned}
		&\Theta_{n^{\prime}} = a_{n^{\prime}}Q_{n^{\prime}n^{\prime}},\\
		&\Theta_{n^{\prime}}^{(1)} = \sum_{m=-\infty}^\infty v_{n^{\prime}m}\text{i}k_1a_m (Q_{n^{\prime}n^{\prime}} - Q_{n^{\prime}m}) - v_{n^{\prime}m}\sqrt{\mu_{n^{\prime}}}a_m (S_{n^{\prime}n^{\prime}} - S_{n^{\prime}m}).
\end{aligned}
\end{equation}
The functions $Q_{nm}$ and $S_{nm}$ in the above equations are given by
\begin{equation}
\begin{aligned}
	Q_{nm} = \sum_{j=0}^1\frac{1}{\kappa_{nj}} \Big( & \frac{1}{\mu_A}\big(\text{e}^{\text{i}\kappa_{nj} \chi_{j+1}}\Gamma(c+\epsilon_{j+1};\mu_m, \mu_A) - \text{e}^{\text{i}\kappa_{nj} \chi_{j}}\Gamma(c+\epsilon_{j};\mu_m, \mu_A)\big)\\
	&  - \frac{1}{\mu_{Bnj}} \text{e}^{\text{i}\kappa_{nj} (\chi_j - (c+\epsilon_j)/\sigma_j)}\big(\Gamma(c+\epsilon_{j+1};\mu_m,\mu_{Bnj}) - \Gamma(c+\epsilon_{j}; \mu_m,\mu_{Bnj})\big) \Big),
	\label{Q_nm}
\end{aligned}
\end{equation}
\begin{equation}
	\begin{aligned}
            S_{nm} = \sum_{j=0}^1 \frac{1}{\text{i}\kappa_{nj}}\Big( & \frac{1}{\sqrt{\eta_{Am}}}\big(\text{e}^{\text{i}\kappa_{nj} \chi_{j+1}}\mathrm{E}(\eta_{Am}(c+\epsilon_{j+1})) - \text{e}^{\text{i}\kappa_{nj} \chi_{j}}\mathrm{E}(\eta_{Am}(c+\epsilon_{j})))\big)\nonumber\\
            &  - \frac{1}{\sqrt{\eta_{Bmj}}} \text{e}^{\text{i}\kappa_{nj} (\chi_j - (c+\epsilon_j)/\sigma_j)}\big(\mathrm{E}(\eta_{Bmj}(c+\epsilon_{j+1})) - \mathrm{E}(\eta_{Bmj}(c+\epsilon_{j}))\big) \Big),
	\end{aligned}
	\label{S_nm}
\end{equation}
where the function $\Gamma$ is defined by
\begin{equation}
    \Gamma(x;\mu,\nu) = \text{e}^{\text{-i}\nu x} \mathrm{E}(\mu x) - \sqrt{\frac{\mu}{\mu-\nu}}\mathrm{E}((\mu - \nu)x) + \frac{1}{1-\text{i}}(1-\text{e}^{\text{-i}\nu x}),
	\label{Gamma}
\end{equation}
and
\begin{equation}
	\begin{aligned}
		&\mu_A  = k_1 + k/\beta^2(M_0-x_1/S_0),\\
		&\mu_{Bnj} = k_1 - (k_{2n} - kx_2/S_0)/\sigma_j,\\
		&\kappa_{nj} = k_{2n} - kx_2/S_0 + k/\beta^2(M_0-x_1/S_0)\sigma_j,\\
                &\eta_{Am} = K_m + kM_0/(\beta^2 + \sigma^2) - k/\beta^2(M_0-x_1/S_0),\\
                &\eta_{Bmj}= K_m + kM_0/(\beta^2 + \sigma^2) + (k_{2n} - kx_2/S_0)/\sigma_j.
	\end{aligned}
	\label{wavenumbers2}
\end{equation}
Note that (\ref{equ:PSD}) is the far-field sound induced by the scattered pressure only. When the incident pressure is also incorporated, as pointed out by \citet{Amiet1978}, the number $1$ appearing in the parenthesis of function $\Gamma$ should be omitted, i.e. the third term on the right hand side of (\ref{Gamma}) should be replaced by  $-\text{e}^{\text{-i}\nu x}/(1-\text{i})$.

\nomenclature[a-pf]{$p_f$}{Far field sound pressure}
\nomenclature[a-N]{$N$}{$(2N+1)$ denoting the number of periods of serrations}
\nomenclature[g-Theta]{$\Theta_{n^{\prime}}$}{Fundamental order of the gust-response function}
\nomenclature[g-Theta]{$\Theta_{n^{\prime}}^{(i)}$}{$i$-th order of the gust-response function}

\nomenclature[g-Gamma]{$\Gamma$}{Properly bounded complex function defined in (\ref{Gamma})}
\nomenclature[a-Q]{$Q_{nm}$}{Function defined for expressing wall pressure gust response}
\nomenclature[g-muA]{$\mu_{A}$}{Wavenumber in $Q_{nm}$}
\nomenclature[g-muB]{$\mu_{Bnj}$}{Wavenumber in $Q_{nm}$}
\nomenclature[g-kappa]{$\kappa_{nj}$}{Wavenumber in $Q_{nm}$}
\nomenclature[a-S]{$S_{nm}$}{Function defined for expressing wall pressure gust response}
\nomenclature[g-etaa]{$\eta_{Am}$}{Wavenumber in $S_{nm}$}
\nomenclature[g-etab]{$\eta_{Bmj}$}{Wavenumber in $S_{nm}$}

\subsubsection{Statistical formulation}
The hypothetical surface pressure of frequency $\omega$ beneath a turbulent boundary layer on the plate surface that would exist when the semi-infinite plate is absent can be expressed as a Fourier integral,
\begin{equation}
	P_{int}(\omega,x^{\prime},y^{\prime}) = \iint P_{i}(\omega, k_1, k_2) \text{e}^{\text{i}(k_1 x^{\prime} + k_2 y^{\prime})} \ud k_1 \ud k_2.
	\label{equ:GustIntegral}
\end{equation}
Generally, for a given frequency $\omega$, $k_1$ can have different values~\citep{Amiet1976}. However, experiments~\citep{Willmarth1959} have shown that $P_{i}(\omega, k_1, k_2)$ peaks in the vicinity of $k_1 = \omega/ U_c$, where the convection velocity $U_c$ is only a weak function of $\omega$. Hence (\ref{equ:GustIntegral}) reduces to
\begin{equation}
	P_{int}(\omega,x^{\prime},y^{\prime}) = \int_{-\infty}^{\infty} P_{i}(\omega, k_2) \text{e}^{\text{i}(k_1 x^{\prime} + k_2 y^{\prime})} \ud k_2.
	\label{equ:GustIntegralSimplified}
\end{equation}

\nomenclature[a-P]{$P_{int}$}{Surface pressure beneath the turbulent boundary layer}

As shown in the preceding section, a wall pressure gust of
$$P_{i}(\omega,k_2)\text{e}^{\text{-i}(\omega t-k_1 x^{\prime}-k_2 y^{\prime})}$$
will induce a far-field sound pressure
\begin{equation}
	\left(\frac{-\text{i}\omega x_3c}{4\pi c_0 S_0^2}\right)\lambda \frac{\sin\big((N+1/2)\lambda(k_2-kx_2/S_0)\big)}{\sin\big(\lambda/2(k_2-kx_2/S_0)\big)}\mathcal{L}(\omega,k_2)P_{i}(\omega,k_2).
	\label{equ:Temp}
\end{equation}
Thus, the wall pressure defined by (\ref{equ:GustIntegralSimplified}) will induce a far-field sound pressure of
\begin{equation}
	p_f(\mathbf{x},\omega) = \left(\frac{-\text{i}\omega x_3 c}{4\pi c_0 S_0^2}\right)\int_{-\infty}^{\infty} \lambda\frac{\sin\big((N+1/2)\lambda(k_2-kx_2/S_0)\big)}{\sin\big(\lambda/2(k_2-kx_2/S_0)\big)}\mathcal{L}(\omega,k_2)P_{i}(\omega,k_2)\ud k_2.
	\label{equ:GustResponseIntegral}
\end{equation}
The PSD of the far-field sound is given by
\begin{equation}
  S_{pp}(\mathbf{x}, \omega) = \lim_{T\to \infty} \left(\frac{\pi}{T}\langle p_f(\mathbf{x},\omega) p_f^{\ast}(\mathbf{x},\omega)\rangle \right),
	\label{equ:DefinitionOfSpp}
\end{equation}
where the asterisk denotes complex conjugate, and $2T$ is the time length used to obtain $p_f(\mathbf{x},\omega)$ by performing Fourier transformation. Substituting (\ref{equ:GustResponseIntegral}) into (\ref{equ:DefinitionOfSpp}) yields
\begin{equation}
		S_{pp}(\mathbf{x},\omega) = \left(\frac{\omega x_3c}{4\pi c_0 S_0^2}\right)^2 \int_{-\infty}^{\infty} \lambda^2\left(\frac{\sin\big((N+1/2)\lambda(k_2-kx_2/S_0)\big)}{\sin\big(\lambda/2(k_2-kx_2/S_0)\big)}\right)^2 \left|\mathcal{L}\right|^2 \Pi(\omega,k_2)\ud k_2,
	\label{equ:SerratedPSD}
\end{equation}
where $\Pi(\omega, k_2)$ is the wavenumber spectral density~\citep{Amiet1975} of the hypothetical wall pressure beneath the turbulent boundary layer on the plate surface. For very wide serrations, i.e. $h \approx 0$, (\ref{equ:SerratedPSD}) reduces to Amiet's model~\citep{Amiet1976}. Equation~(\ref{equ:SerratedPSD}) can be simplified by assuming a very large span, i.e. the number of serrations $(2N + 1)$ is sufficiently large. Using the following equation,
\begin{equation}
	\lim_{N\to \infty}{\lambda^2\frac{\sin^2\big((N+1/2)\lambda(k_2-kx_2/S_0)\big)}{\sin^2\big(\lambda/2(k_2-kx_2/S_0)\big)}} \sim 2\pi d \sum_{m=-\infty}^{\infty}\delta(k_2-kx_2/S_0+2m\pi/\lambda),
	\label{equ:DeltaSimplification}
\end{equation}
where $\delta(x)$ is the conventional generalized function defined in Sec.~\ref{sec:II}, one can show that the PSD of the far-field sound in the plane $y^{\prime} = 0$ is given by
\begin{equation}
  S_{pp}(\mathbf{x}, \omega) = \left(\frac{\omega x_3c}{4\pi c_0 S_0^2}\right)^2 2\pi d \sum_{m=-\infty}^{\infty}\left|\mathcal{L}(\omega,2m\pi /\lambda) \right|^2\Pi(\omega,2m \pi/\lambda).
        \label{equ:FundamentalResult}
\end{equation}

\nomenclature[z-PSD]{PSD}{Power Spectral Density}
\nomenclature[g-Pi]{$\Pi$}{Wavenumber spectral density}
\nomenclature[g-delta]{$\delta$}{Conventional generalized function in section~\ref{sec:FloquetAndFourier} and (\ref{equ:DeltaSimplification})}

Equation~\ref{equ:FundamentalResult} is the fundamental result of this paper and it is interesting to note that the infinite series in (\ref{equ:FundamentalResult}) appears similar to that in Howe's model shown in (\ref{equ:NondimensionalForm}). For example, both results show that the PSD of far-field sound is related to the wavenumber spectral density of the surface pressure through $\Pi(\omega,2m\pi/\lambda)$, therefore a skewed wall pressure gust with $k_2 = 2 m \pi/\lambda$ plays an important role in sound generation.

\subsection{Discussion on the effects of serration geometry}
\label {subsec:Discussion}
The complicated formulation of the far-field noise (\ref{equ:FundamentalResult}), and the response function (\ref{equ:SerratedResponseFunction}), make it very difficult to assess the effectiveness of serrations without numerical evaluation of the equations. This section attempts to derive two simple conditions for serrations to obtain effective noise reduction. In order to achieve significant sound reduction, we wish to minimize (\ref{equ:FundamentalResult}). Since $|\mathcal{L}(\omega, 2m\pi/\lambda)|^2$ is very complex, we will perform an order analysis first.

Careful examination of (\ref{equ:SerratedResponseFunction}) shows that $|\mathcal{L}|^2$ is proportional to $1/|\kappa_{n^\prime j}|^2$. For illustration purposes, we assume that the observer is at $90^\circ$ above the trailing-edge in the mid-span plane, i.e. $x_1 = 0$ and $x_2 = 0$, and that the Mach number is low, e.g. $M_0 < 0.2$. Then $\kappa_{mj} \approx 0$ when $m$ satisfies $k_2 + 2m\pi/\lambda \approx 0$ and thus the value of $|\mathcal{L}(\omega, -2m\pi/\lambda)|^2$ is dominated by mode $m$, i.e.
\begin{equation}
  |\mathcal{L}(\omega, -2m\pi/\lambda)|^2 \approx \frac{2}{\lambda^2 c^2} \left|a_{m} Q_{mm} + \Theta_{m}^{(1)} + \Theta_{m}^{(2)} + \cdots\right|^2.
    \label{equ:SimplifiedL}
\end{equation}
Furthermore, noting that $v_{nm} = 4h/(\pi^2(n- m)^2)$ when $n-m$ is odd, equation (\ref{eq:Thetan'}) suggests that $\Theta_{m}^{(i)}$ may roughly be approximated by only summing over modes $m-1$ and $m+1$, since higher orders $m \pm (2j + 1)$ with $j \geq 1$ are at least one order of magnitude smaller due to the quadratic term in the denominator of $v_{nm}$. Using this approximation, $\Theta_{m}^{(i)}$ varies linearly with $a_{m-1}$ and $a_{m+1}$. It can be shown from (\ref{equ:anexpression}) that $|a_{m-1}|$ and $|a_{m+1}|$ are of the order of $|a_{m}|$ and $|\Theta_{m}^{(i)}|=O(|a_m|)$. From~(\ref{equ:SimplifiedL}), we hence have $|\mathcal L(\omega, -2m\pi/\lambda)|^2=O(|a_m|^2)$, and
\begin{equation}
  \sum_{m = -\infty}^{\infty}|\mathcal{L}(\omega, 2m\pi/\lambda)|^2 \Pi(\omega, 2m\pi/\lambda) = O\left(\sum_{m = -\infty}^{\infty} |a_m|^2 \Pi(\omega, 2m\pi/\lambda)\right).
  \label{equ:orderEquation}
\end{equation}

We are now in a position to discuss the conditions for minimizing (\ref{equ:orderEquation}).
 From the definition of $a_m$ in (\ref{an}), it can be shown that
\begin{equation}
  a_m=\frac{\text{e}^{\text{i}m\pi/2}}{2}\text{sinc}(k_1h - m\pi/2) +\frac{\text{e}^{-\text{i}m\pi/2}}{2}\text{sinc}(k_1h + m\pi/2).
  \label{equ:anexpression}
\end{equation}
Thus, $|a_m|$ is maximum when $m \approx \pm \nu_0$, where $\nu_0 = 2k_1h/\pi$. To minimize the right hand side in~(\ref{equ:orderEquation}), we therefore require that $\Pi(\omega, 2m\pi/\lambda) \ll \Pi(\omega, 0)$ when $m$ approaches $\pm\nu_0$.
% Apparently, the largest value of $|a_m|$ (the order of $O(1)$) is % obtained when $k_1 h \pm m\pi/2 \approx 0$, and for $k_1 h \pm m\pi/2 % \neq 0$
% \begin{equation}
%   |a_m| \le  \frac{1}{k_1h \pm m\pi/2}.
%   \label{equ:anexpressionorder}
% \end{equation}
% where the sign before $m\pi/2$ on the right hand side of % (\ref{equ:anexpression}) is determined to minimise $k_1h \pm m\pi/2$. % As $|a_m|$ obtains a large value when $m \approx \pm 2k_1h/\pi$, in % order to obtain a small value of (\ref{equ:orderEquation}), we wish % $\Pi(\omega, 2m\pi/\lambda) \ll \Pi(\omega, 0)$ for $|m| \ge N_0$, % where $N_0 \approx 2k_1 h/\pi$.
Assuming frozen turbulence, $\Pi(\omega, 2m\pi/\lambda)$ is given by
\begin{equation}
  \Pi \left(\omega,\frac{2\pi m}{\lambda}\right) =\frac{1}{2\pi} \int_{-\infty}^{\infty} S_{qq}(\omega,y^\prime)\text{e}^{-\text{i}\frac{2\pi m}{\lambda}y^\prime} \ud y^\prime.
    \label{equ:Pi}
\end{equation}
When $2\pi\nu_0 l_{y'}/\lambda  = k_1 l_{y'} \sigma \gg 1$, the integrand in (\ref{equ:Pi}) for $m$ close to (or larger than) $\nu_0$ oscillates rapidly within the length scale $l_{y'}$ of
$S_{qq}(\omega, y^\prime)$, which corresponds to the spanwise
correlation length given by
\begin{equation}
    l_{y^\prime}(\omega) = \frac{1}{S_{qq}(\omega,0)}\int_{-\infty}^{\infty} S_{qq}(\omega,y^\prime) \ud y^\prime.
    \label{equ:ly}
\end{equation}
The integral in~(\ref{equ:Pi}) therefore evaluates to a small value compared to $\Pi(\omega, 0)$. Thus, a condition for noise reduction is that $k_1 h_e \gg 1$, where we have defined an effective root-to-tip length $2h_e = \sigma l_{y^\prime}$ that describes the correlated serration height.

% Thus, for $|m| > N_0$, the condition $k_1 l_{y^\prime}(\omega) \sigma % \gg 1$ implies $\Pi(\omega, 2m\pi/\lambda)
% \ll \Pi(\omega, 0)$. By analogy with the definition of the root-to-tip % length $2h = \sigma \lambda/2$, we define an effective root-to-tip % length $2h_e = \sigma l_{y^\prime}(\omega)$, which corresponds to the % correlated serration height. Far-field trailing-edge noise can hence % be effectively reduced when $k_1 h_e \gg 1$.
\nomenclature[a-h]{$h_e$}{Half of the equivalent root-to-tip length}

As $2k_1 h_e = k_1 \sigma l_{y^\prime}(\omega)$, the decay rate of spanwise correlation length $l_{y^\prime}(\omega)$ with respect to frequency is critical. If one makes use of Corcos's correlation length model~\citep{Corcos1964}, $l_{y^\prime}(\omega) \approx 2.1 U_c / \omega$, $k_1 h_e$ will reduce to a constant, in this case $2.1\sigma$, a sole function of the serration sharpness factor and independent of frequency. This is consistent with the findings of \citet{Howe1991}. However, if the decay rate were faster than that given by Corcos, no sound reduction or even some sound increase would occur at high frequencies. An accurate description of the characteristics of the surface pressure fluctuation beneath a boundary layer is therefore critical for the model to accurately predict the sound reduction at high frequencies.

% It is worth noting that $k_1 h_e \gg 1$ does not necessarily guarantee % (\ref{equ:orderEquation}) obtains a smaller value than that of a % straight trailing-edge. Since when $k_1 h \to 0$, one finds that $N_0 % \to 0$ and (\ref{equ:orderEquation}) reduces to the same order of % magnitude as Amiet's model for straight % trailing-edge~\citep{Amiet1976}, hence no sound reduction can be % achieved. In another words, $k_1 h_e \gg 1$ is only a necessary % condition for achieving significant trailing-edge noise reduction. % When $k_1 h$ is sufficiently large and $k_1 h_e \gg 1$,
% \begin{equation}
%   \sum_{m = -\infty}^{\infty}|\mathcal{L}(\omega, 2m\pi/\lambda)|^2 %   \Pi(\omega, 2m\pi/\lambda) \sim \sum_{m = -N_0+1}^{N_0-1} %   \frac{1}{(k_1h \mp m\pi/2)^2}\Pi(\omega, 2m\pi/\lambda).
%   \label{equ:finalOrder}
% \end{equation}
% By noting the quadratic term in (\ref{equ:finalOrder}) and that % $\Pi(\omega, k_2)$ decreases when $k_2$ increase, we expect larger % values of $k_1h$ yield smaller values of (\ref{equ:finalOrder}). Thus, % the second necessary condition for trailing-edge noise reduction is % that $k_1 h \gg 1$. This same criterion was first derived % in~\cite{Howe1991, Howe1991a}.

Note that the condition $k_1 h_e \gg 1$ is only a necessary condition, because when $k_1 h \rightarrow 0$ there is no noise reduction. This can be seen from equation (\ref{equ:anexpression}), since $a_m=0$ when $k_1 h \rightarrow 0$ except when $m=0$, so the right hand side in equation~(\ref{equ:orderEquation}) reduces to $\Pi(\omega, 0)$ which corresponds to the straight edge case. From~(\ref{equ:anexpression}), for a given integer $m$ away from $\nu_0$, $|a_m|$ tends to zero when $k_1h \gg 1$. This provides another necessary condition for noise reduction. Physically, the height of the serrations must be sufficient for them to be seen by the incoming hydrodynamic waves.
% Since the edge is tilted relative to the incoming waves, the phase of % the scattered field will vary along the edge of the serrations. If the % serration height is sufficiently large, these phase differences will % lead to destructive interference and noise reduction. This phenomenon % is discussed in more detail in section~\ref{sec:V}.

\nomenclature[g-N]{$\nu_0$}{$2k_1h/\pi$}
We have thus obtained two necessary conditions for noise reduction, $k_1h_e \gg 1$ and $k_1 h \gg 1$, that are consistent with those proposed by Howe. These conditions will be further investigated in section~\ref{sec:V}.

\section{Results}
\label{sec:III}
\subsection{Model Validation}
\subsubsection{FEM implementation}
\label{sec:FEM}
For the coupled differential equations mentioned in the last section, the solutions are obtained by performing an iterative-solving procedure. In this section, we shall investigate the validity of the proposed iterative solution using the Finite Element Method (FEM). Instead of solving the far-field sound directly, costing a significant amount of computer memory, a feasible alternative is to calculate the near field using FEM and obtain the far-field solution by performing a surface integral, as adopted in analytical models, see~(\ref{equ:SerratedFarField}).

In order to make a direct comparison between the computational and analytical results, the wave equation together with the boundary conditions given in (\ref{equ:TimeWaveEquation}) and (\ref{equ:BoundaryConditionsForSerrated}), respectively, will be solved. The governing equation and boundary conditions in the frequency domain can be written as,
\begin{equation}
	\left\{
		\begin{aligned}
			&\beta^2 \frac{\partial^2 P}{\partial x^{\prime 2}} + \frac{\partial^2 P}{\partial y^{\prime 2}}+ \frac{\partial^2 P}{\partial z^{\prime 2}} + 2ikM_0\frac{\partial P}{\partial x^{\prime}} + k^2 P = 0,\\
			& \frac{\partial P}{\partial z^{\prime}}(x^{\prime},y^{\prime},0) = 0, &x^{\prime}< H(y^{\prime})\\
			& P(x^{\prime},y^{\prime},0) = -P_{i}\text{e}^{\text{i}( k_1x^{\prime} + k_2 y^{\prime})}, & x^{\prime} \ge H(y{^\prime}).
		\end{aligned}
		\right.
		\label{equ:ComputedEquations}
\end{equation}
Using the transformation $P = \bar{P} \text{e}^{-\text{i}kM_0/\beta^2 x^\prime}$, the first-order derivative term, induced by the background flow, can be eliminated. The results will be transformed back to the physical domain before making comparisons. The scattered near-field pressure is obtained by solving (\ref{equ:ComputedEquations}) using FEM and the far-field sound pressure is obtained by integrating the pressure distribution over surface, as described in (\ref{equ:SerratedFarField}) and (\ref{equ:PhaseRelation}).

\begin{figure}
	\centering
	\includegraphics[width=0.55\textwidth]{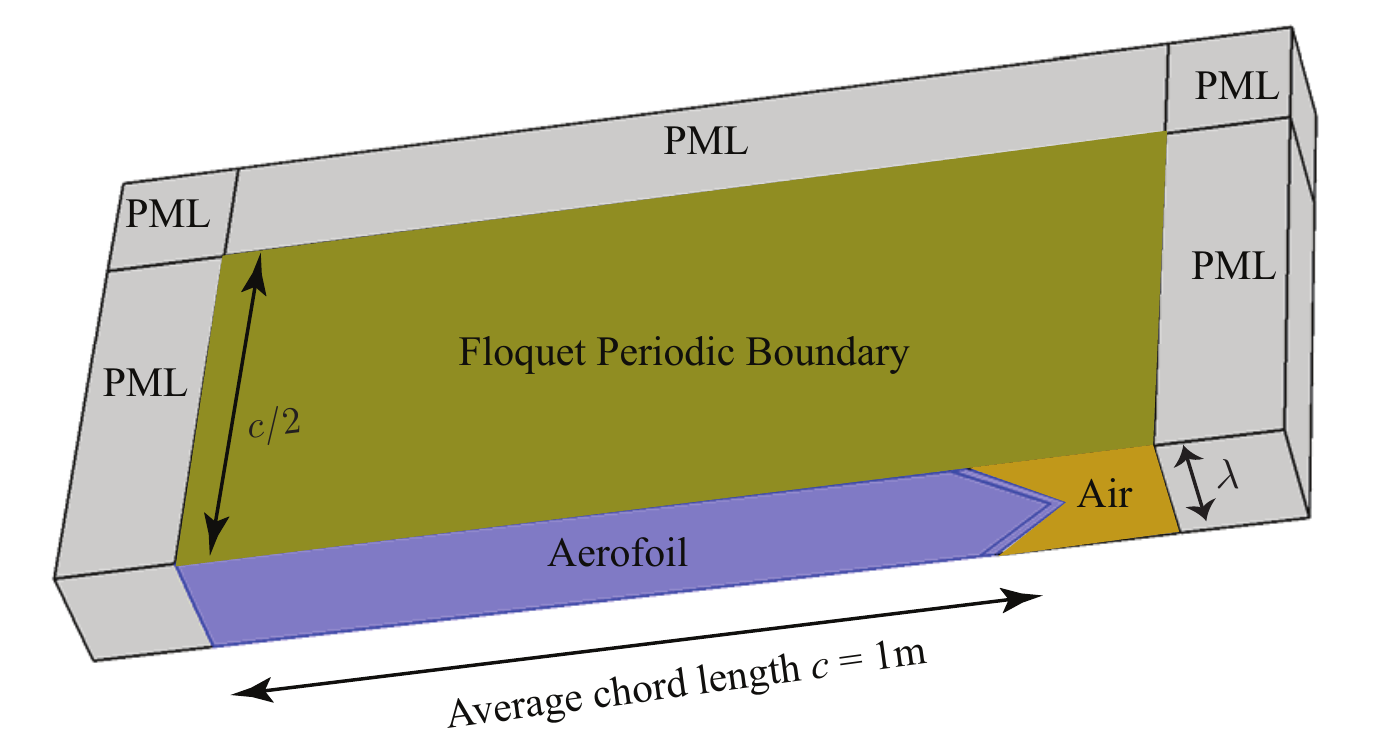}
	\caption{Illustration of the FEM computing domain and the boundary conditions}
	\label{fig:ComputingDomain}
\end{figure}

The commercial software COMSOL 4.4 is used to perform the FEM simulations. Simulations are performed for a single serration and chord length $c = 1m$, as shown in figure~\ref{fig:ComputingDomain}. For the boundary conditions, the normal velocity on the surface of the plate vanishes, while the pressure values are fixed over the wake half plane. In addition, as illustrated in figure~\ref{fig:ComputingDomain}, the walls on both the upper and lower sides of computational domain represent Floquet periodic boundary conditions. The radiation boundary condition is implemented via Perfectly Matched Layers (PMLs), as shown in figure~\ref{fig:ComputingDomain}. The mesh is made of tetrahedron cells with quadratic shape functions. The mesh is highly non-uniform and is generated to accurately resolve the hydrodynamic pressure fluctuations near the serrated edge and the acoustic pressure perturbation in the far-field. A mesh sensitivity test has been carried out to ensure the proper convergence of the simulation. In the hydrodynamic region, the mesh contains more than 10 grid points within one hydrodynamic wavelength. The same ratio of 10 grid points per wavelength is used in the far-field, relative to the acoustic wavelength. The final mesh contains about $0.6$ million elements.
\begin{figure}
	\centering
	\includegraphics[width=0.55\textwidth]{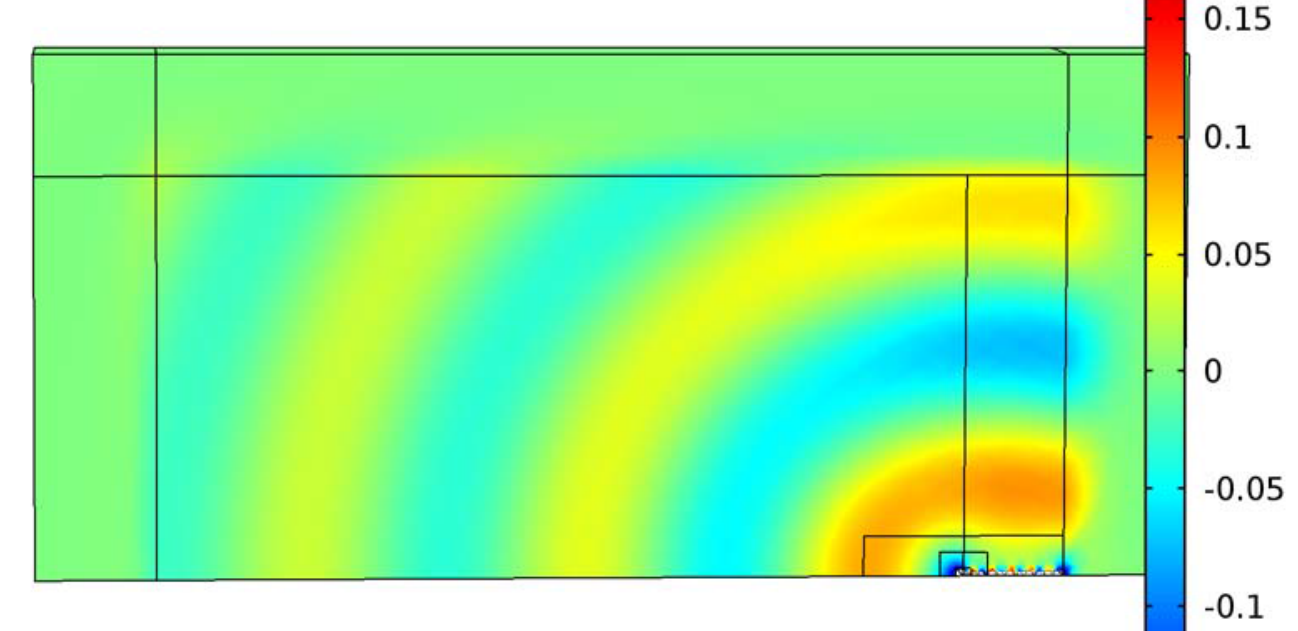}
	\caption{Computed pressure distribution scattered by a straight trailing-edge }
	\label{fig:StraightEdgeComputation}
\end{figure}

\nomenclature[z-PML]{PML}{Perfectly Matched Layer}

Figure~\ref{fig:StraightEdgeComputation} shows the results for a straight trailing-edge at $f = 1000$~Hz and $M = 0.1$. The turbulent convection velocity is assumed to be $U_c = 0.7 U$, where $U = M_0 c_0$. The wavenumbers are  $k_1 = 2\pi f/U_c$, $k_2 = 0$ and the amplitude of incident wall pressure gust $P_{i}$ is unity. It can be seen that the PMLs do not cause spurious reflections. The solution was also compared with Amiet's analytical solution. The maximum difference at $f=1000$ Hz was less than $1\%$. The error is partly due to the infinite-chord assumption in the analytical model and in part to the FEM numerical errors.

\subsubsection{FEM model validation}
\begin{figure}
	\centering
	\subfigure[$M_0  = 0.1,\,\lambda/h=6,\,h/c=0.025$]{\label{fig:Validation155Mach01}\includegraphics[width=0.49\textwidth]{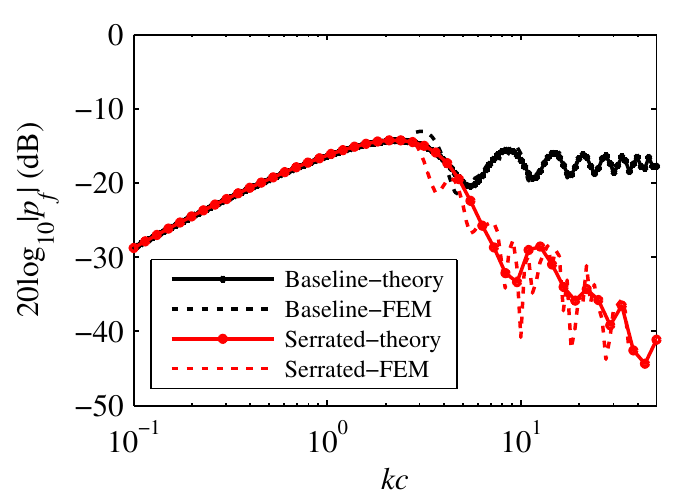}}
	\subfigure[$M_0 = 0.2,\,\lambda/h=6, \, h/c = 0.025$]{\label{fig:Validation155Mach02}\includegraphics[width=0.49\textwidth]{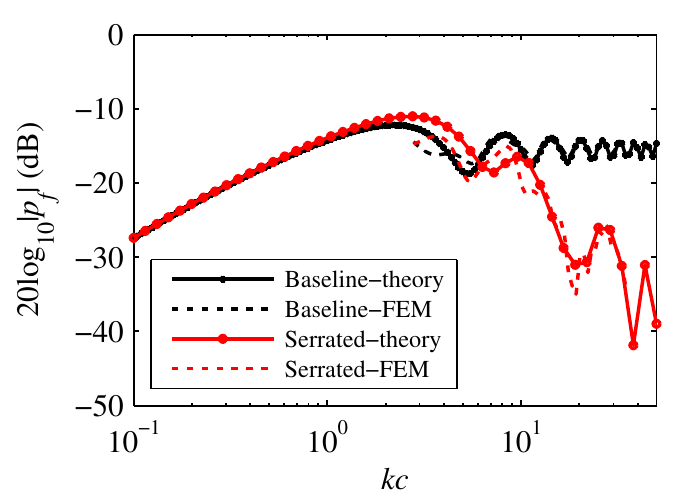}}
	\subfigure[$M_0 = 0.1, \,\lambda/h=3,\, h/c =0.05$]{\label{fig:Validation1510Mach01}\includegraphics[width=0.49\textwidth]{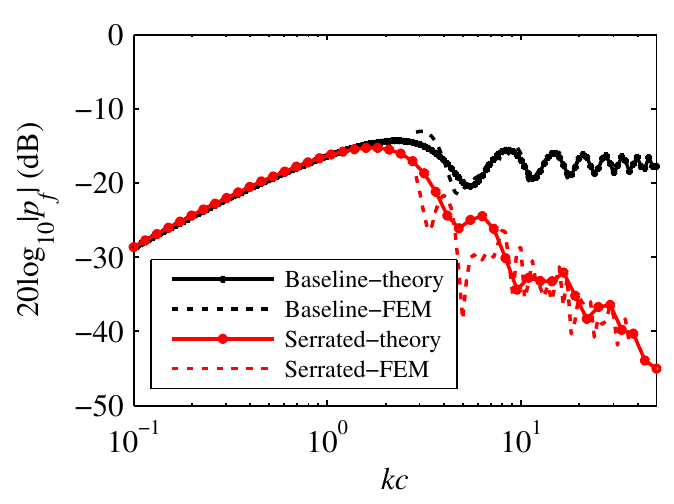}}
	\subfigure[$M_0 = 0.2,\,\lambda/h=3,\, h/c =0.05$]{\label{fig:Validation1510Mach02}\includegraphics[width=0.49\textwidth]{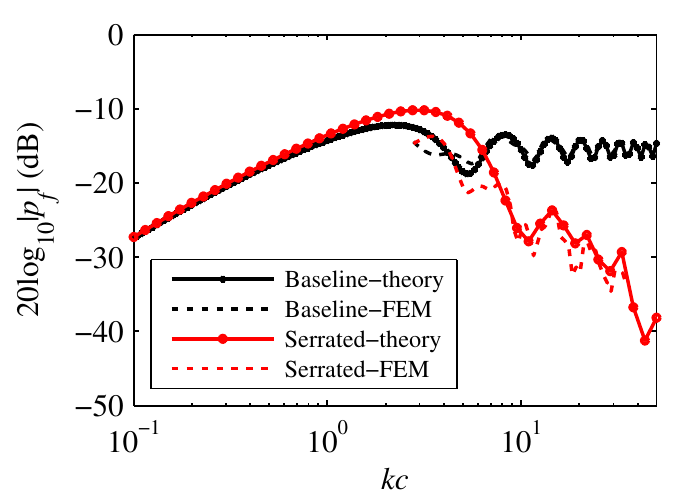}}
	\subfigure[$M_0 = 0.2,\, \lambda/h=0.5, \,h/c = 0.1 $]{\label{fig:Validation520Mach02}\includegraphics[width=0.49\textwidth]{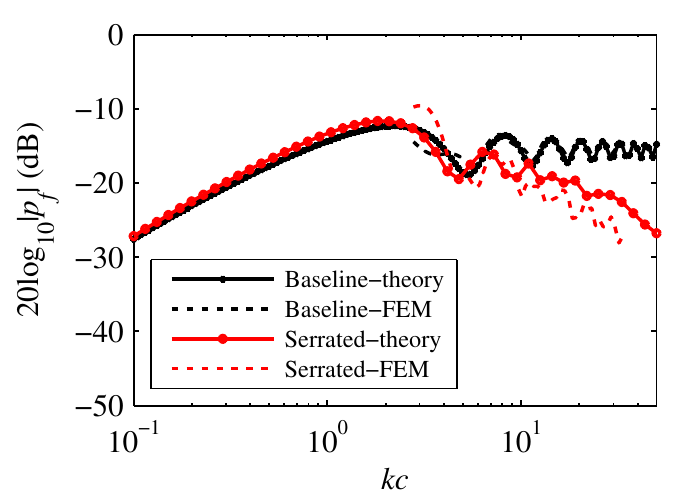}}
	\subfigure[$M_0 = 0.2,\, \lambda/h=0.3, \,h/c = 0.1$]{\label{fig:Validation320Mach02}\includegraphics[width=0.49\textwidth]{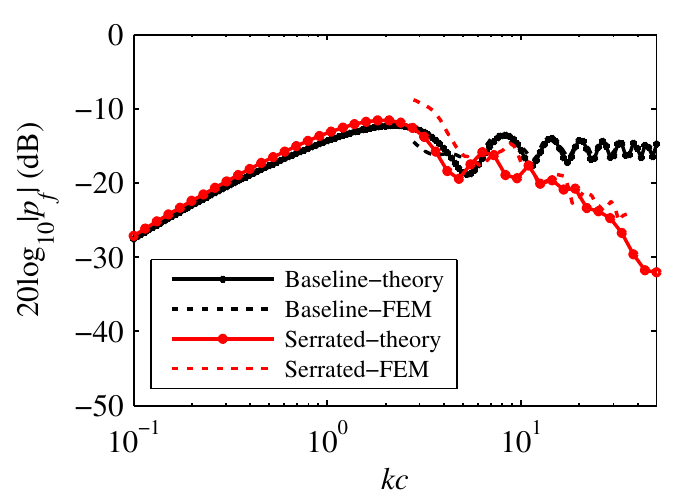}}
		\caption{SPL at $90^\circ$ above the trailing-edge in the mid-span plane with $x_3 = 1$ due to a wall pressure gust of frequency $\omega$ with $k_2 =0$}
\end{figure}

This section provides a comprehensive comparison between the analytical model for gust-induced far-field noise developed in Sec.~\ref{subsubsec:far-fieldPressure} and the FEM model developed in Sec.~\ref{sec:FEM} for different serrations and at different Mach numbers. The far-field sound pressure induced by a wall pressure gust of $k_2 = 0$ at different frequencies is chosen for comparison.  Note that in the mathematical model, we use the second-order approximation of the gust-response function $\mathcal{L}$, as shown in (\ref{equ:SerratedResponseFunction}). The chord length $c$ is chosen to be unity with a span-to-chord ratio of 8. The Mach number and geometrical shape of the serrations can vary in different cases.

Results are presented for a range of serrations, as illustrated in figure~\ref{fig:Validation155Mach01} through \ref{fig:Validation320Mach02}. In this study, we shall only focus on low Mach numbers, i.e. $M_0 \le 0.2$. The far-field pressure, obtained from the FEM model at $90^\circ$ above the trailing-edge in the mid-span plane with $x_3 =1$, is plotted as $20\log_{10}{|p_f(\mathbf{x},\omega)|}$ against the theoretical predictions. Results are provided for both the baseline (no serration, i.e. $h = 0$) and serrated trailing-edges. It can be found that a sound reduction of more than 20 dB can be achieved for the far-field sound induced by this specific wall pressure gust. This reduction should however not to be confused with the sound reduction of the real trailing-edge noise, which comprises different wall pressure gusts at different values of $k_2$.

Figures~\ref{fig:Validation155Mach01} to \ref{fig:Validation1510Mach02} show very good agreement between the theoretical and computational results. The low frequency discrepancies originate from the infinite chord assumption in the analytical model. The serration cases presented in figures~\ref{fig:Validation155Mach01} to \ref{fig:Validation1510Mach02} are not normally considered sharp enough, based on experimental observations ~\citep{Gruber2012} to reduce the noise significantly. Figures~\ref{fig:Validation520Mach02} and \ref{fig:Validation320Mach02} show the results at $M_0= 0.2$ for serrations with $\lambda/h = 0.5$ and $h/c = 0.1$ and $\lambda/h=0.3$ and $h/c = 0.1$, respectively. It can be seen from these two figures that for sharper serrations, the average error between the numerical calculations and theoretical predictions normally increases, which might be caused by the relatively slower convergence rate of the second-order approximations of (\ref{equ:SerratedResponseFunction}) compared to that for the wide serrations. The agreement between the FEM results and the proposed model, however, is generally good, suggesting that the second-order solution does indeed give a reasonable good approximation for (\ref{equ:SerratedResponseFunction}). The issue of the convergence of the iterative method will also be discussed later.

\subsection{The far-field sound spectrum}
\label{subsec:Results}
A parametric study of far-field noise reduction was carried out by Howe~\citep{Howe1991, Howe1991a}, indicating the possibility of significant noise reduction, much higher than measured data~\citep{Gruber2012}. In this section, we shall use the second-order iterative model developed in Sec.~\ref{sec:II} and carry out a parametric study. For illustration purposes, we adopt Chase's model~\citep{Chase1987} of the wavenumber spectral density. It is argued by \citet{Chase1987} that the convection velocity $U_c$ is weakly dependent on frequency and on average $U_c \approx 0.7U$. According to Chase's model, the wavenumber spectral density is well approximated by
\begin{equation}
\Pi(\omega, k_1, k_2) =	\frac{C_m \rho_0^2 v_{\ast}^3k_1^2 \delta^5}{\left( (k_1 - \omega/U_c)^2(\delta U_c/3v_{\ast})^2 + (k_1^2+k_2^2)\delta^2 + \chi^2\right)^{5/2}},
	\label{equ:ChaseModel}
\end{equation}
where $\rho_0$ is the density of air, and $C_m \approx 0.1553, \chi \approx 1.33, v_{\ast} \approx 0.03U$. The turbulent boundary layer thickness $\delta$ in (\ref{equ:ChaseModel}) is approximated by~\citep{Eckert1959}
\begin{equation}
  \delta/c = 0.382Re_c^{-1/5},
  \label{equ:bounaryLayerThickness}
\end{equation}
where $Re_c$ is the Reynolds number based on chord $c$. An inspection of (\ref{equ:ChaseModel}) shows that the wavenumber spectrum peaks around $k_1 = \omega/U_c$. Here, we assume $k_1  = \omega/U_c$ and obtain $\Pi(\omega,k_2)$ by integrating (\ref{equ:ChaseModel}) with respect to $k_1$ and then keeping the leading order terms~\citep{Howe1991}, which yields
\begin{equation}
	\Pi(\omega,k_2) \approx \frac{4C_m \rho_0^2 v_{\ast}^4(\omega/U_c)^2 \delta^4}{U_c\left( ((\omega/U_c)^2+k_2^2)\delta^2 + \chi^2\right)^{2}}.
	\label{equ:IntegratedChaseModel}
\end{equation}
Substituting (\ref{equ:IntegratedChaseModel}) into (\ref{equ:FundamentalResult}) and using $(\rho_0 v_{\ast}^2)^2 (d/c_0)$~\citep{Howe1991} to non-dimensionalize the far field PSD yields
\begin{equation}
    \frac{S_{pp}(\mathbf{x},\omega)}{(\rho_0 v_{\ast}^2)^2 (d/c_0)} = \frac{C_m}{2\pi}\Psi(\mathbf{x},\omega),
	\label{equ:NormalizedSpp}
\end{equation}
where $\Psi(\mathbf{x},\omega)$ is defined as
\begin{equation}
\Psi(\mathbf{x},\omega) = \left(\frac{x_3c}{S_0^2}\right)^2\left(\frac{U_c}{c_0}\right) \sum_{m=-\infty}^{\infty}\left|\mathcal{L}(\omega,2\pi m/\lambda) \right|^2 \frac{(\omega\delta/U_c)^4}{\left[(\omega\delta/U_c)^2+(2m\pi\delta/\lambda)^2 + \chi^2\right]^{2}}.
	\label{equ:IllustrationEquation}
\end{equation}
\begin{figure}
	\centering
	\subfigure[$\lambda/h=8, h/c=0.025$]{\label{fig:Discussion205Mach01}\includegraphics[width=0.49\textwidth]{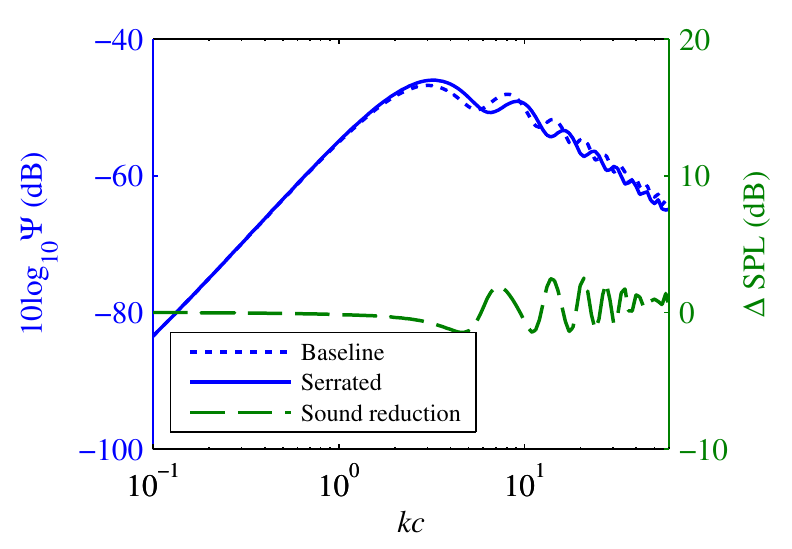}}
	\subfigure[$\lambda/h=4, h/c = 0.025$]{\label{fig:Discussion2010Mach01}\includegraphics[width=0.49\textwidth]{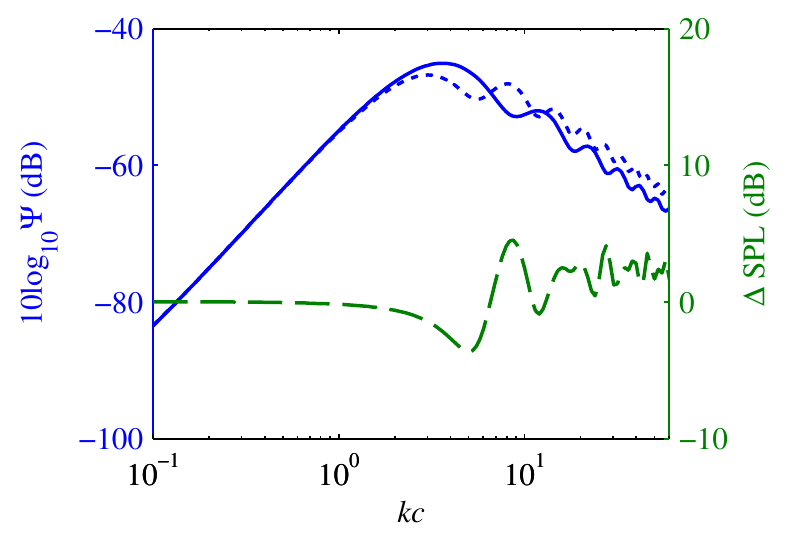}}
	\subfigure[$\lambda/h=2, h/c = 0.05$]{\label{fig:Discussion1010Mach01}\includegraphics[width=0.49\textwidth]{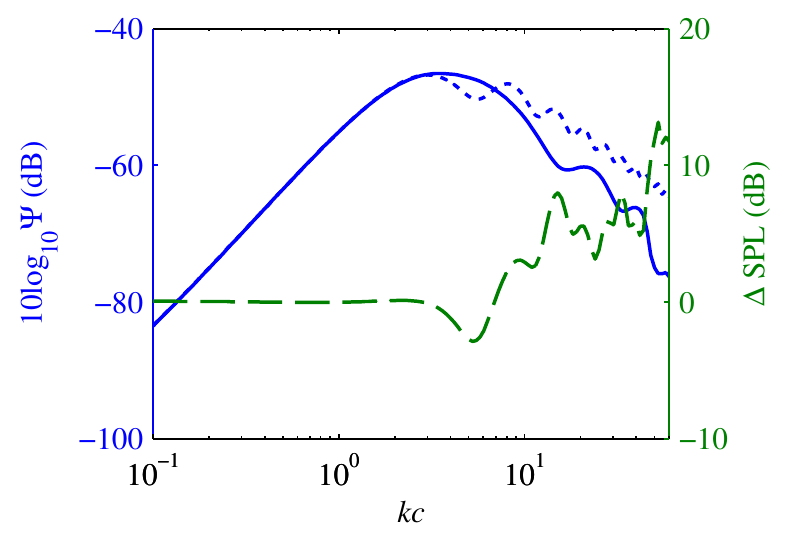}}
	\subfigure[$\lambda/h=1, h/c = 0.05$]{\label{fig:Discussion510Mach01}\includegraphics[width=0.49\textwidth]{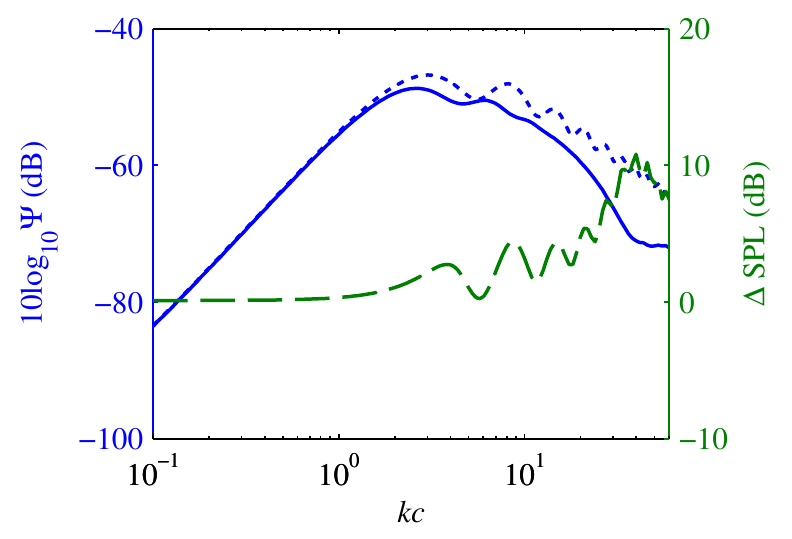}}
	\subfigure[$\lambda/h=0.4,h/c = 0.05 $]{\label{fig:Discussion210Mach01}\includegraphics[width=0.49\textwidth]{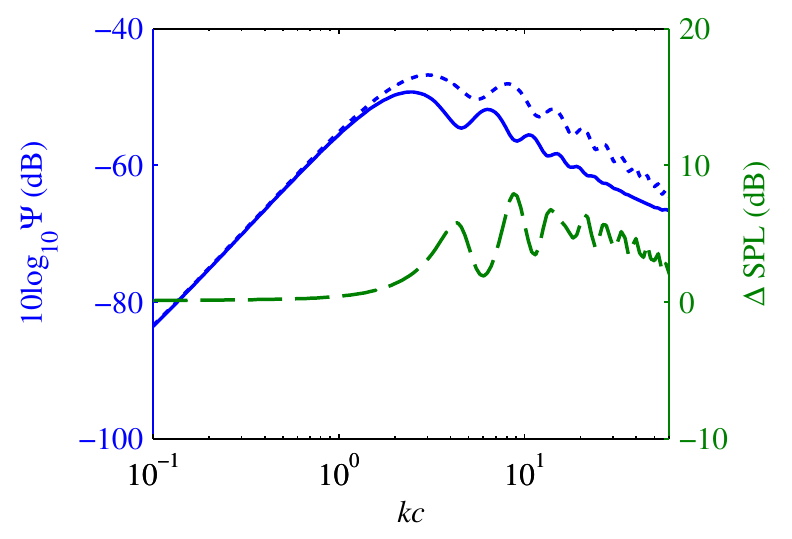}}
	\subfigure[$\lambda/h=0.2, h/c = 0.05$]{\label{fig:Discussion110Mach01}\includegraphics[width=0.49\textwidth]{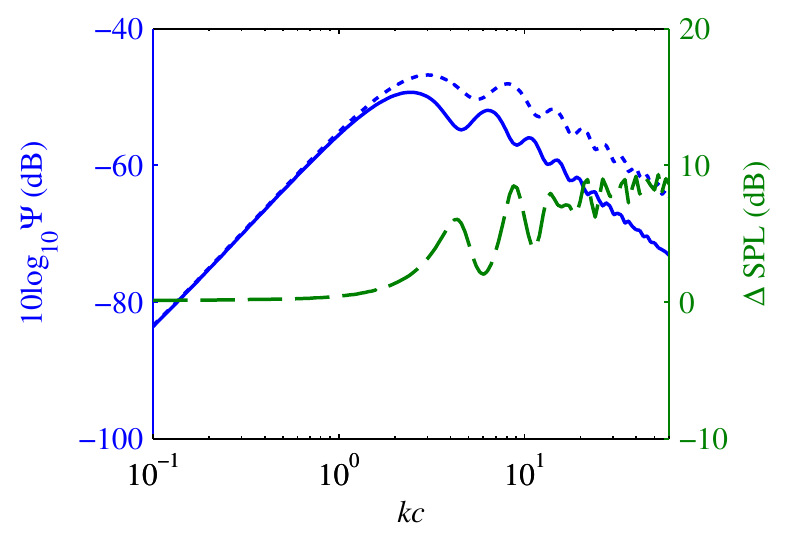}}
        \caption{The normalized spectrum of equation~\ref{equ:IllustrationEquation} (left axis) for straight and serrated trailing-edges, in solid and short dashed blue lines respectively, and the sound reduction spectrum $\Delta$SPL (right axis) with $M_0= 0.1$ and an observer at $90^\circ$ above the trailing-edge in the mid-span plane with $x_3 = 1$.}
	\label{fig:Discussion2052010Mach01}
\end{figure}

\nomenclature[a-Cm]{$C_m$}{Dimensionless constant in Chase's model}
\nomenclature[g-chi]{$\chi$}{Dimensionless constant in Chase's model}
\nomenclature[a-vast]{$v_{\ast}$}{Friction velocity in Chase's model}
\nomenclature[g-rho0]{$\rho_0$}{Density of air}
\nomenclature[g-delta]{$\delta$}{Boundary layer thickness}
\nomenclature[x-par]{$[\text{ }]$}{The nearest integer function}
\nomenclature[a-Mbar]{$\bar{M}$}{Integer number denoting decay rate of $\Pi(\omega,k_2$)}

\nomenclature[g-Psi]{$\Psi$}{Normalized power spectrum of far-field sound}
\nomenclature[a-M]{$M$}{Mode number where $\lvert a_n\rvert$ peaks}

The following figures are plotted using (\ref{equ:IllustrationEquation}) for a variety of serration geometrical parameters. Since much of the experimental work focused on the trailing-edge noise at low Mach numbers~\citep{Gruber2013}, we shall only focus on the low Mach numbers, i.e. $M_0 \le 0.2$. The function $\mathcal{L}$ in (\ref{equ:IllustrationEquation}) is defined in (\ref{equ:SerratedResponseFunction}) and we take the second-order approximation here. Note that the incident pressure is also taken into consideration~\citep{Amiet1978}. The observer point is at $90^\circ$ above the trailing-edge in the mid-span plane, namely $(x_1= 0,\, x_2 =0,\, x_3 = 1)$. It is worth pointing out that in figure~\ref{fig:Discussion2052010Mach01} both the far-field sound spectrum and the sound reduction spectrum are shown.

The normalized sound power spectrum at $M_0 = 0.1$ for different serrations are shown in figures~\ref{fig:Discussion205Mach01} to \ref{fig:Discussion110Mach01}. The spectrum for a serrated trailing-edge with $4h/\lambda = 0.5$ is shown in figure~\ref{fig:Discussion205Mach01}. As expected, the sound reduction is approximately zero over the entire the frequency range of interest. Increasing the sharpness of the serrations gradually improves the sound reduction performance, as shown in figures~\ref{fig:Discussion2010Mach01}, \ref{fig:Discussion1010Mach01} and \ref{fig:Discussion510Mach01}. For sufficiently sharp serrations, significant sound reduction is achieved over a wide range of frequencies, as shown in figure~\ref{fig:Discussion210Mach01}, where the sharpness factor is $4h/\lambda = 10$. The result obtained for a sawtooth serration with $\lambda/h = 0.2, h/c = 0.05$ at $M_0 = 0.1$ is shown in figure~\ref{fig:Discussion110Mach01}. Comparing figures~\ref{fig:Discussion210Mach01} and~\ref{fig:Discussion110Mach01} suggests that for already sharp serrations, further increasing the sharpness can provide a better high frequency noise reduction performance while the low frequency performance ($kc < 10$) remains unchanged. For the sharp serrations presented in figure~\ref{fig:Discussion110Mach01} the far-field sound is reduced by about 10~dB at high frequencies. This better agrees with experiments where a noise reduction of up to $7{-10}$ dB is observed~\citep{Dassen1996, Parchen1999}. From figure~\ref{fig:Discussion205Mach01} and \ref{fig:Discussion2010Mach01} it can be found that a slight noise increase may occur at low frequencies. In fact, the noise increase becomes even more pronounced at low frequencies when the Mach number is high, e.g. $M_0 = 0.4$. The explanation of the noise increase at low frequencies will be given in Sec.~\ref{sec:V}.
\begin{figure}
  \centering
	\subfigure[$\lambda/h= 2,h/c = 0.025 $]{\label{fig:convergenceWide}\includegraphics[width=0.49\textwidth]{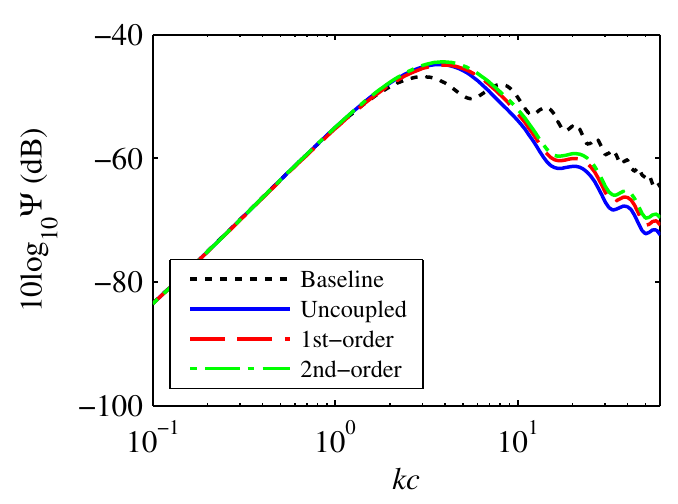}}
	\subfigure[$\lambda/h= 0.2,h/c = 0.05 $]{\label{fig:convergenceNarrow}\includegraphics[width=0.49\textwidth]{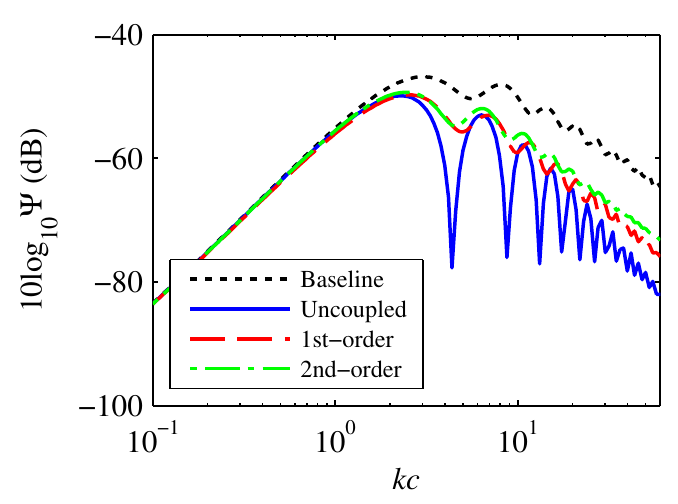}}
  \caption{The normalized spectrum for straight and serrated trailing-edges obtained using different order approximations, $M_0= 0.1$, the observer is at $90^\circ$ above the trailing-edge in the mid-span plane with $x_3 = 1$.}
  \label{fig:convergence}
\end{figure}

The results presented in figure~\ref{fig:Discussion2052010Mach01} were based on the second-order approximation. The convergence rate of different order solutions can be inspected by presenting the far-field sound spectrum using different-order approximations, as shown in figure~\ref{fig:convergence}, where the far-field spectrum using zero, first and second-order approximations are presented. Figure~\ref{fig:convergence}(a) presents results for a wide serration with $\lambda/h = 2$. As expected, due to the weak coupling between different modes the first and second-order solutions yield almost the same results. It is thus safe to assume that the second-order approximation gives an accurate solution for wide serrations. Figure~\ref{fig:convergence}(b) shows the convergence results for a narrow serrations with $\lambda/h = 0.2$. It can be seen that the difference between the first and second-order solutions is much smaller than that between the zero- and first-order ones. The maximum difference between the first and second-order approximations at high frequencies is less than 2 decibels. Thus, the second-order solution can be assumed to provide a reasonably accurate solution for narrow serrations even at high frequencies.

\subsection{Directivity patterns}
\begin{figure}
	\centering
	\subfigure[$kc = 1$]{\label{fig:Mach1kc1}\includegraphics[width=0.4\textwidth]{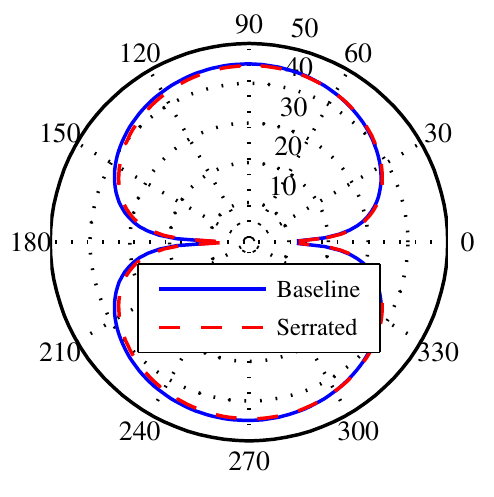}}
	\subfigure[$kc = 3$]{\label{fig:Mach1kc3}\includegraphics[width=0.4\textwidth]{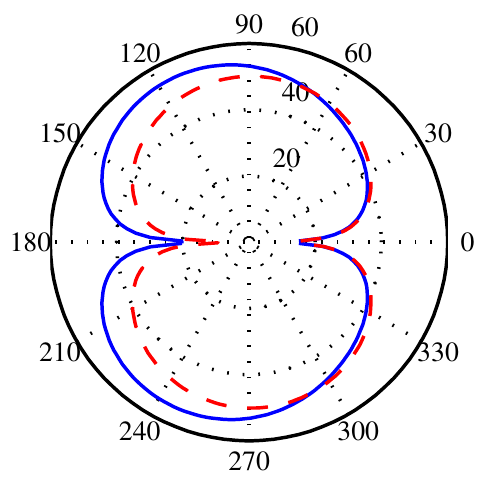}}
	\subfigure[$kc = 5$]{\label{fig:Mach1kc5}\includegraphics[width=0.4\textwidth]{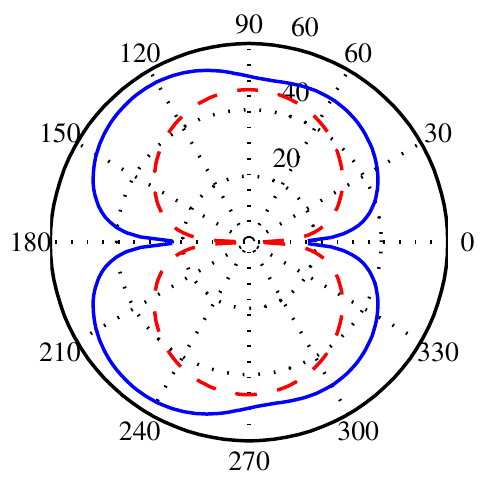}}
	\subfigure[$kc = 10$]{\label{fig:Mach1kc10}\includegraphics[width=0.4\textwidth]{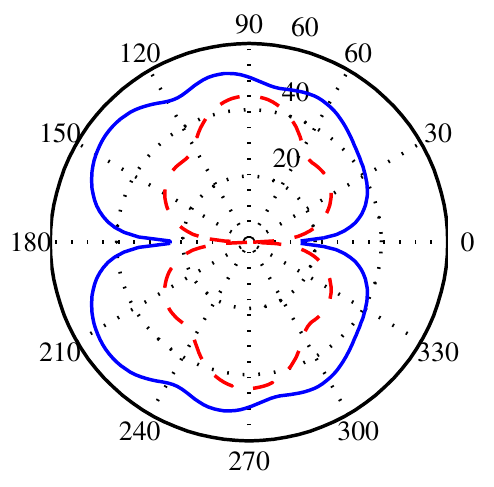}}
	\subfigure[$kc = 20$]{\label{fig:Mach1kc20}\includegraphics[width=0.4\textwidth]{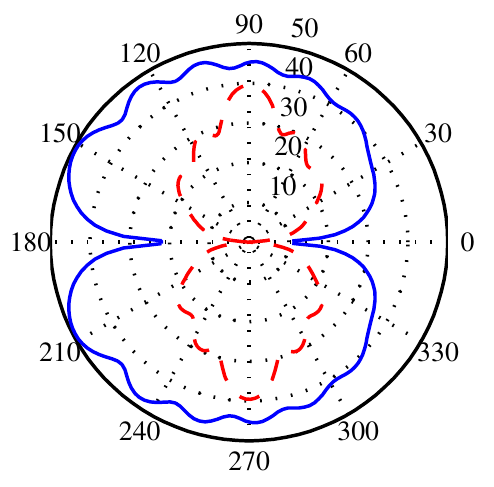}}
	\subfigure[$kc = 50$]{\label{fig:Mach1kc50}\includegraphics[width=0.4\textwidth]{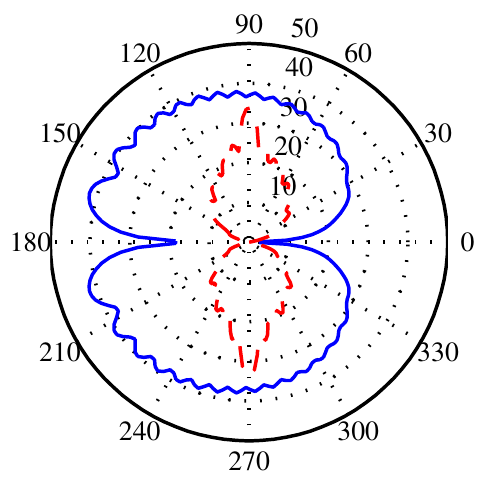}}
        \caption{The directivity patterns plotted against $\theta$ in the mid-span plane ($x_2 = 0$ and $r = \sqrt{(x_1^2 + x_3^2)} = 1$) at $M_0 = 0.1$ for serrations with $\lambda/h = 0.4$ and $h/c = 0.05$.  The far-field sound level in the figures are $10\log_{10} \left(\Psi(\mathbf{x}, \omega)/4\cdot10^{-10}\right)$.}
	\label{fig:Mach1Directivity}
\end{figure}

\begin{figure}
	\centering
	\subfigure[$kc = 1$]{\label{fig:Mach4kc1}\includegraphics[width=0.4\textwidth]{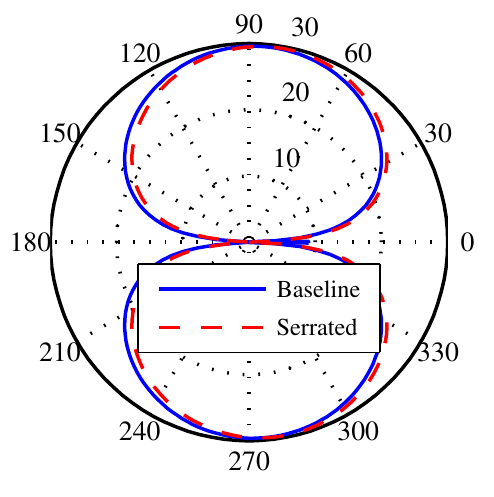}}
	\subfigure[$kc = 3$]{\label{fig:Mach4kc3}\includegraphics[width=0.4\textwidth]{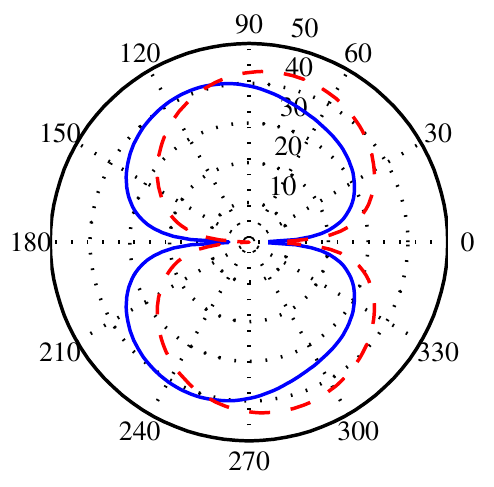}}
	\subfigure[$kc = 5$]{\label{fig:Mach4kc5}\includegraphics[width=0.4\textwidth]{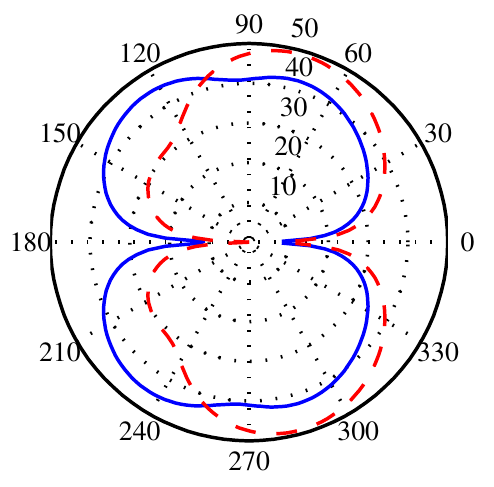}}
	\subfigure[$kc = 10$]{\label{fig:Mach4kc10}\includegraphics[width=0.4\textwidth]{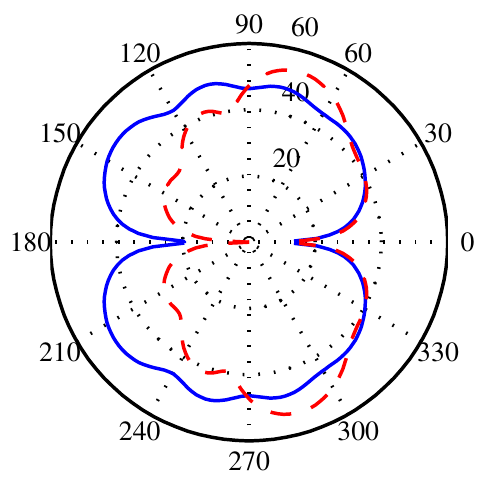}}
	\subfigure[$kc = 20$]{\label{fig:Mach4kc20}\includegraphics[width=0.4\textwidth]{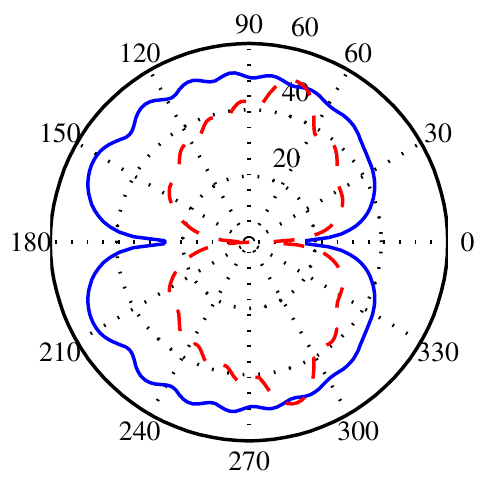}}
	\subfigure[$kc = 50$]{\label{fig:Mach4kc50}\includegraphics[width=0.4\textwidth]{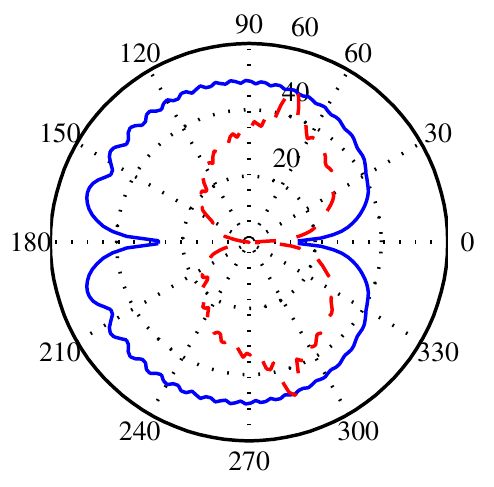}}
        \caption{The directivity patterns plotted against $\theta$ in the mid-span plane ($x_2 = 0$ and $r = \sqrt{(x_1^2 + x_3^2)} = 1$) at $M_0 = 0.4$ for serrations with $\lambda/h = 0.4$ and $h/c = 0.05$. The far-field sound level in the figures are $10\log_{10} \left(\Psi(\mathbf{x}, \omega)/4\cdot10^{-10}\right)$.}
	\label{fig:Mach4Directivity}
\end{figure}

It is a well established fact that the trailing-edge noise directivity changes with frequency~\citep{Williams1970, Gruber2013}. However, the effect of serrations on trailing-edge noise directivity has received very little research attention. Figures~\ref{fig:Mach1Directivity} and~\ref{fig:Mach4Directivity} present the non-dimensional far-field PSD, (\ref{equ:IllustrationEquation}), based on the second-order solution.  Results are presented for straight and serrated trailing-edges with $\lambda/h=0.4$ at $M_0=0.1$ and $0.4$, respectively. As expected, using serrations has little effect on noise generation mechanism at very low frequencies, $kc \le 1$. Results, however, show that the serrations can effectively reduce the noise at higher frequencies. As discussed earlier, this is believed to be primarily due to the destructive scattering interference effects. While most experimental investigations~\citep{Gruber2013} have focused on the capability of serrations for reducing the noise at small angles and 90 degrees above the trailing-edge, results in figures~\ref{fig:Mach1Directivity} and~\ref{fig:Mach4Directivity} clearly show that serrations are more effective in reducing the noise at large radiation angles, i.e. towards the leading-edge, $\theta>90^\circ$.  This is a very interesting result as noise measurement in laboratory environment is often limited to 30 to 120 degrees due to the anechoic chamber room size constraint or reflection by the contraction nozzle, etc~\citep{Gruber2012, Moreau2013}.

Results have also shown that the use of serrations can lead to significant changes to the directivity pattern of the scattered pressure field at high frequencies. While one would expect a cardioid pattern for straight edges at high frequencies, associated with the edge scattering of a half-plane~\citep{Williams1970}, results for serrated trailing-edges show that the directivity pattern is more dipolar with a clear peak at a specific angle which depends on both the serration sharpness and the Mach number. Numerical study of the directivity pattern for different serrations has shown that the expected cardioid shape gradually changes to a more dipolar shape as the serration sharpness increases and the directivity peak also gradually moves downstream, towards the trailing-edge. The dipolar behaviour of the noise from serrated trailing-edges means that the leading-edge region, $\theta = 180^\circ$, is much quieter than that for straight trailing-edge. Increasing the Mach number also appears to move the peak angle towards the trailing-edge, as can be seen by comparing figures~\ref{fig:Mach1Directivity} and~\ref{fig:Mach4Directivity}. It is also worth mentioning that in the case of high Mach numbers, see figure~\ref{fig:Mach4Directivity}, the use of serrations can lead to considerable noise increase in the trailing-edge region ($ 0^\circ < \theta < 90^\circ$), for intermediate frequencies, $1<kc<10$.

\section{Comparison with Howe's model}
\label{sec:IV}
The mathematical model and serration geometrical criteria developed by Howe have long been used as a tool to evaluate the effectiveness of trailing-edge serrations and estimate the level of noise reduction~\citep{Gruber2012, Azarpeyvand2013, Jone2012}. However, it has repeatedly been shown that Howe's model overpredicts the level of noise reduction~\citep{Dassen1996, Parchen1999, Gruber2012}. To simplify the model, Howe assumes that the Mach number is sufficiently low to neglect the convection effect, the frequency are sufficiently high, satisfying $\omega h/U_c \gg 1$, the statistical property of the turbulence inside the boundary layer remains the same before and after passing the trailing-edge and the diffraction model is based on the Green's function for straight trailing-edges and the slender-wing approximations.

With the introduction of Chase's surface pressure wavenumber spectral density Model, Howe~\citep{Howe1991a} shows that the far-field PSD is given by
\begin{equation}
	\begin{aligned}
		\frac{S_{pp}(\omega,\mathbf{x})}{(\rho_0 v_{\ast}^2)^2 (d/c_0)(\delta/|\mathbf{x}|)^2} = C_m/\pi \sin^2(\theta/2)\sin(\phi)\Psi_f(\omega),\\
	\label{equ:HowesModelWithChaseModelRepeated}
\end{aligned}
\end{equation}
where
\begin{equation}
		\Psi_f(\omega) = 8 (h/\delta)^2 \sum_{m =-\infty}^{\infty} \frac{(\omega h/U_c)^2[(\omega h/U_c)^2 + (2m\pi h/\lambda)^2][1-\cos(2\omega h/U_c)/\cos(m\pi)]}{[(2 \omega h/U_c)^2 - m^2 \pi^2]^2 [(\omega h/U_c)^2 + (2m\pi  h/\lambda)^2 + (\chi h/\delta)^2]^2}.
	\label{equ:NondimensionalForm}
\end{equation}

\nomenclature[g-theta]{$\theta$}{Fly-over angle, figure~\ref{fig:SerratedFlatPlate}}
\nomenclature[g-phi]{$\phi$}{Angle to the $y^\prime$ axis, figure~\ref{fig:SerratedFlatPlate}}
\nomenclature[g-Psi]{$\Psi_f$}{Normalized power spectrum in Howe's model in (\ref{equ:HowesModelWithChaseModelRepeated})}

Even though the assumption of frozen turbulence is used in both models, Howe's model differs from the model presented in this paper in several ways. In Howe's model, the far-field sound pressure is based on a compact Green's function. The Green's function is obtained by making use of the slender wing approximation. The model developed in this paper, however, gives the scattered sound by solving the convected wave equation. Howe's model neglects the effects of convection, so it is only valid at low Mach numbers. The new model is valid for any subsonic Mach number, as the convection effects have been incorporated in the convected wave equation. In addition, Howe's model requires the high frequency condition $\omega h/U \gg 1$, while the new model requires $kc > 1$, as an infinite chord is assumed in the derivation.
\begin{figure}
    \centering
    \includegraphics{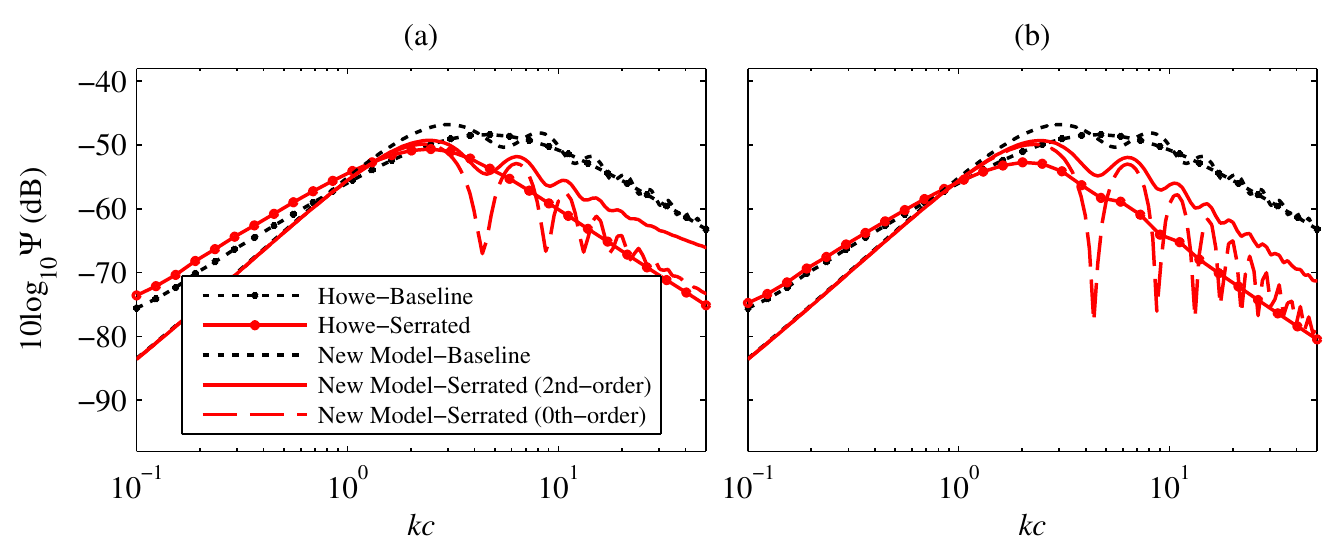}
    \caption{The normalized spectrum of Howe's model and the new model, the observer is at $90^\circ$ above the trailing-edge in the mid-span plane with $x_3 = 1$. (a) $\lambda/h = 0.4,\ h/c = 0.05,\ M_0 = 0.1$; (b) $\lambda/h = 0.2,\ h/c = 0.05, M_0 = 0.1$.}
    \label{fig:ComparisonHowe}
\end{figure}

Figures in this section represent the results obtained using Howe's model, i.e. (\ref{equ:HowesModelWithChaseModelRepeated}), and the model developed in this paper, i.e. (\ref{equ:NormalizedSpp}). The normalized spectrum $\Psi(\mathbf{x},\omega)$ is defined in the same way as in (\ref{equ:NormalizedSpp}). The result of the new model with the correction applied to the third term in (\ref{Gamma}), according to \citet{Amiet1978}, systematically increases the noise level by $6$ dB. Figure~\ref{fig:ComparisonHowe}(a) shows the noise prediction results for a serrated trailing-edge with $\lambda/h = 0.4 , h/c = 0.05$ at $M_0 = 0.1$ using both models. The comparison of the results shows a clear difference between the two methods. At high frequencies, e.g. $kc \approx 50$, Howe's model gives a sound reduction of about $13$ dB while the new model predicts about $7$ dB of noise reduction. Figure~\ref{fig:ComparisonHowe}(b) presents the comparison for shaper serrations, with $\lambda/h = 0.2, h/c = 0.05$ at $M_0 = 0.1$. At $kc \approx 50$, the noise reduction predicted by Howe's model and the new model are, respectively, 18 dB and 10 dB. It can be interpreted from the results that the new model provides a much more realistic noise reduction estimate and is more consistent with experimental observations~\citep{Dassen1996, Parchen1999, Gruber2012, Gruber2013}.

It is very interesting to note that the zero-order solution accurately follows Howe's solution at high frequencies, as shown in figure~\ref{fig:ComparisonHowe}. At intermediate frequencies, i.e. $2 < kc < 20$, the zero-order solution oscillates strongly, but the mean value seems to be following Howe's result. This is actually not hard to understood, as the Green's function used in Howe's model~\citep{Howe1991a}, is in fact only valid locally. In other words, it does not include the coupling effect between adjacent sawtooth edges. Thus, at high frequency we expect the zero-order (without the coupling effect induced by the singular root and tip points) solution coincides with Howe's results. The high-order solution, however adds the coupled interactions between different modes, and this coupling effect clearly reduces the sound reduction predicted by the zero-order solution at high frequencies. Thus, the large overprediction of Howe's model is likely to have been caused by the choice of the Green's function since the Green's function is not able to take into account the coupling effects.

\section{Noise reduction mechanism}
\label{sec:V}
\begin{figure}
    \centering
    \includegraphics[]{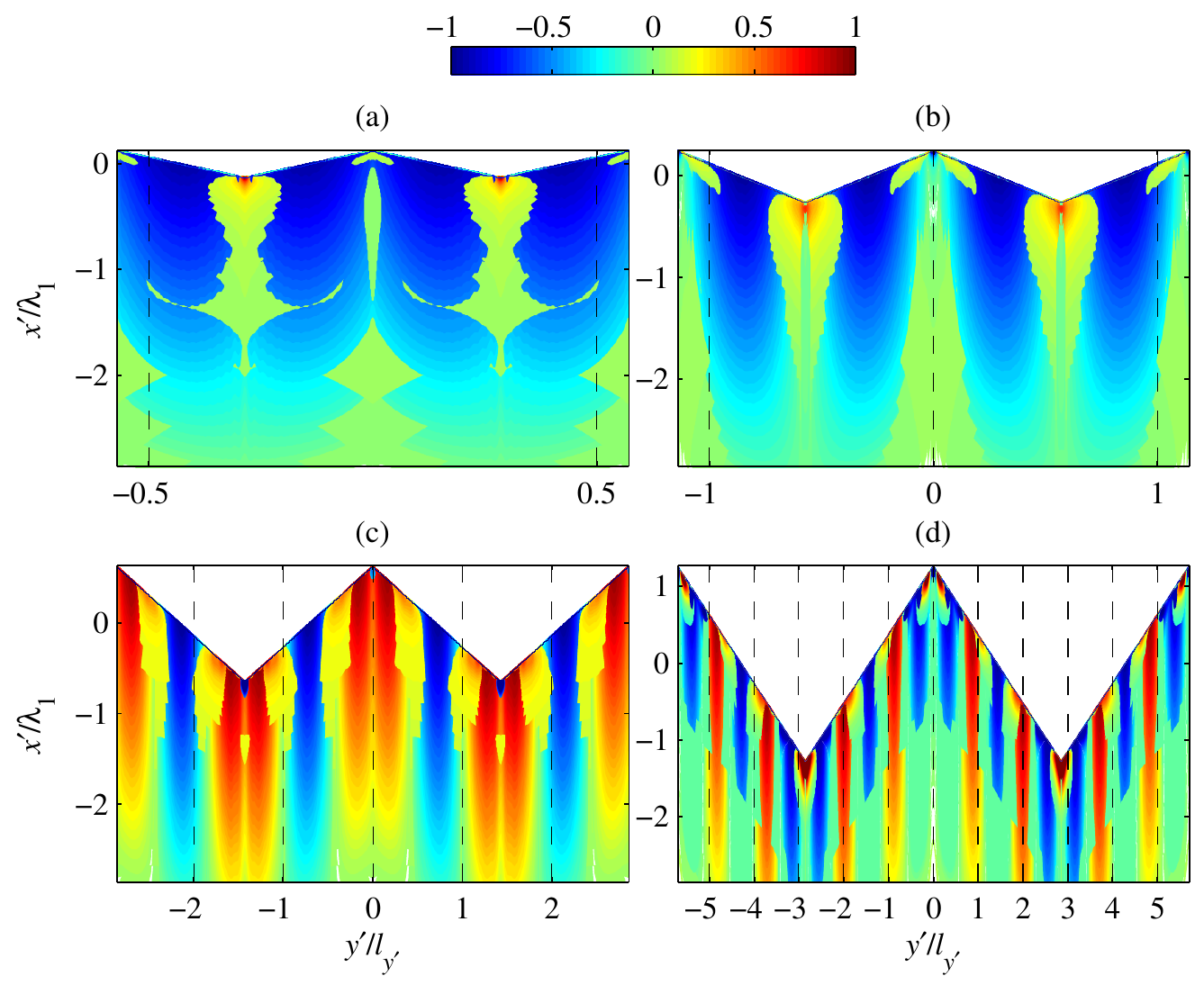}
    \caption{The scattered surface pressure distribution at a fixed frequency for the same $k_1 h_e = 7 $. (a) $k_1 h= 2$; (b) $k_1 h = 4$; (c) $k_1 h = 10$; (d) $k_1 h = 20$.}
    \label{fig:surfacePressureSerratedFreq}
\end{figure}

\begin{figure}
    \centering
    \includegraphics[]{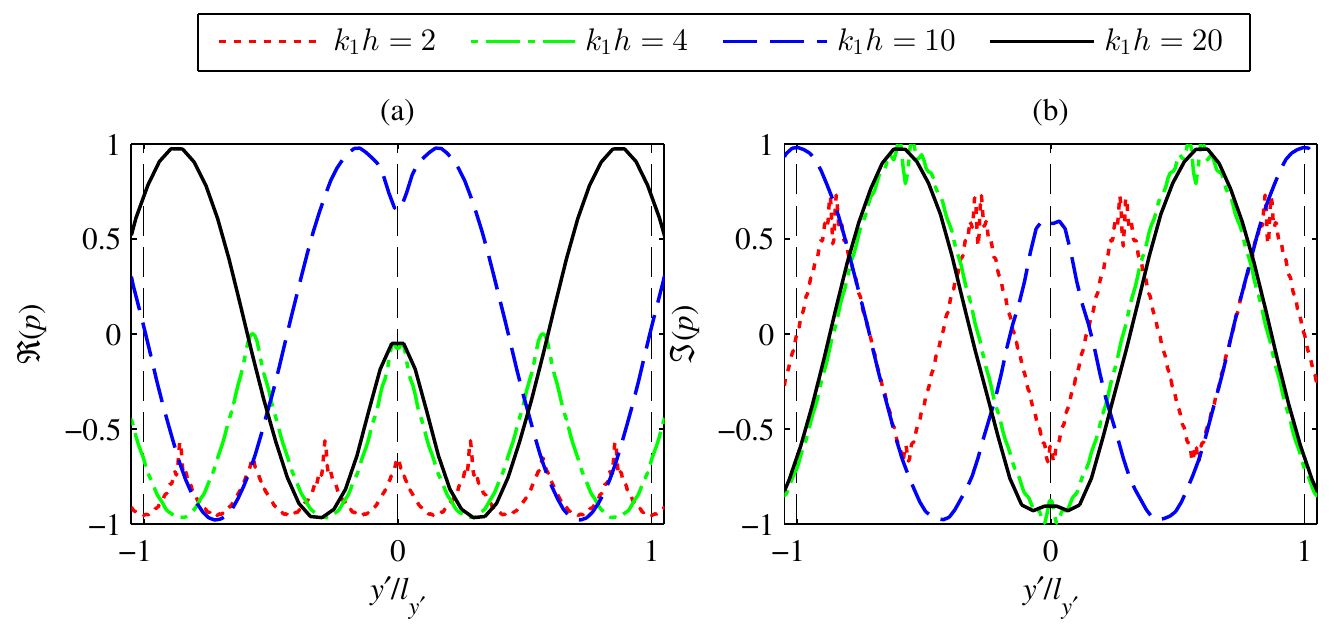}
    \caption{Scattered pressure on the serrated edge. (a) real part; (b) imaginary part.}
    \label{fig:pressureOnEdgeFrequency}
\end{figure}

\begin{figure}
    \centering
    \includegraphics[]{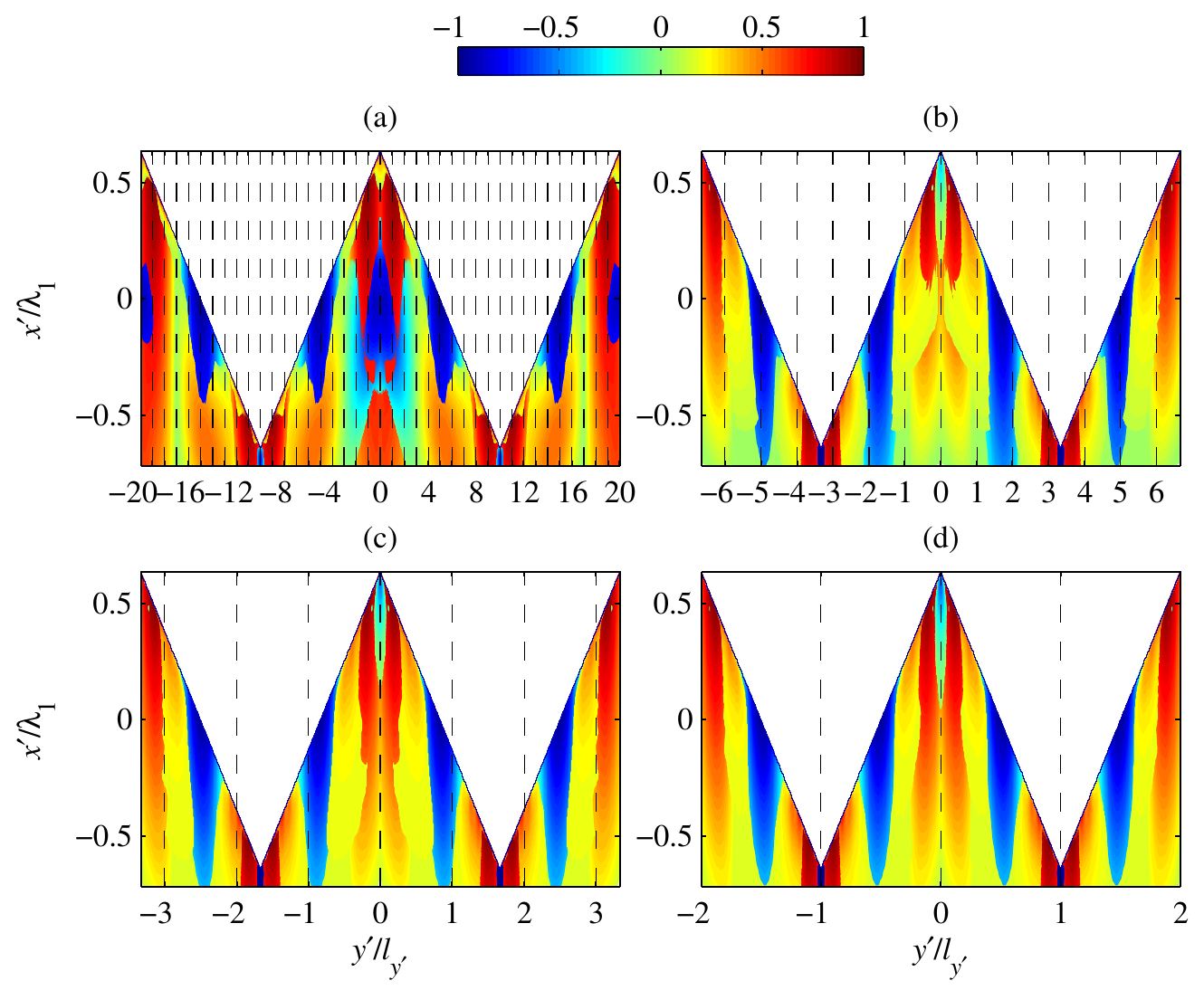}
    \caption{The scattered surface pressure distribution for different serrations at a fixed frequency and $k_1 h = 10$ (a) $k_1 h_e = 1$; (b) $k_1 h_e = 3$; (c) $k_1 h_e = 6$; (d) $k_1 h_e = 10$.}
    \label{fig:surfacePressureSerratedSharpness}
\end{figure}

\begin{figure}
    \centering
    \includegraphics[]{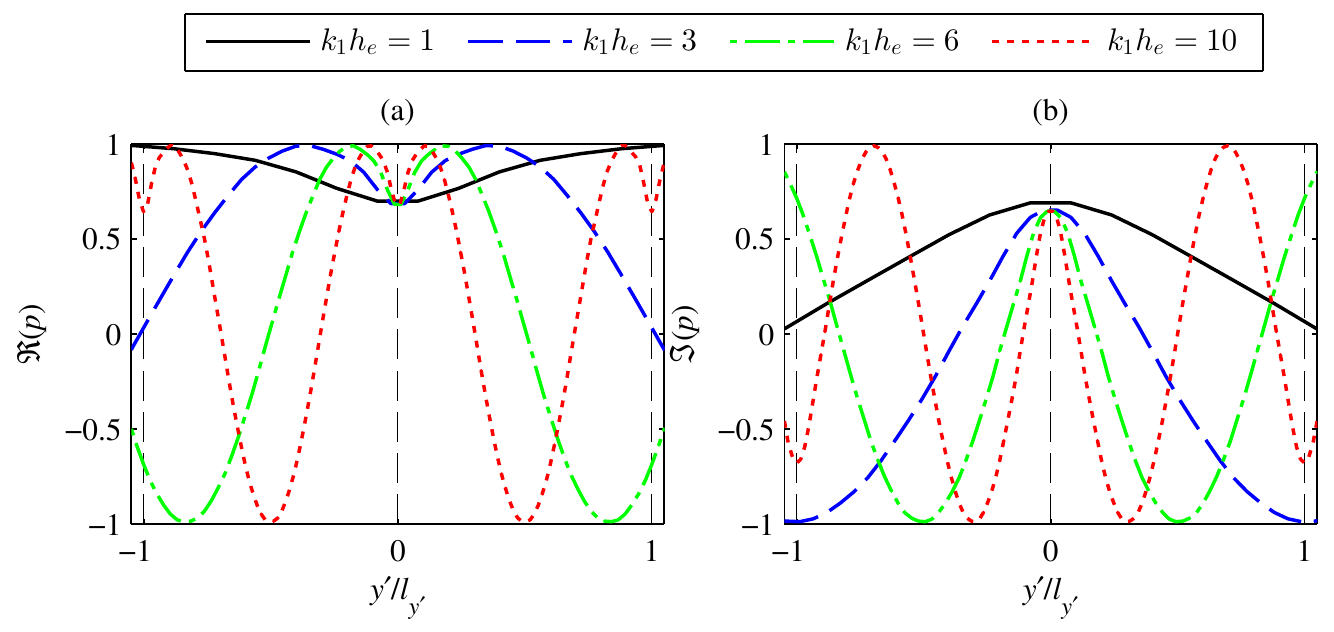}
    \caption{Scattered pressure on the serrated edge for different serration geometries. (a) real part; (b) imaginary part. }
    \label{fig:pressureOnEdgeSharpness}
\end{figure}

In order to better understand the noise reduction mechanism, the pressure distribution over the flat plate surface is presented, see figures~\ref{fig:surfacePressureSerratedFreq} to \ref{fig:pressureOnEdgeSharpness}. As mentioned earlier, the incident pressure only raises the far-field sound by 6 dB systematically, thus it suffices to consider the scattered pressure distribution only. The scattered pressure, as mentioned in Sec.~\ref{sec:II}, is essentially the pressure jump across the flat plate. As mentioned in Sec.~\ref{subsec:Discussion}, the two non-dimensional parameters $k_1h$ and $k_1 h_e$ play an important role for effective sound reduction using serrated trailing-edges. In what follows, the scattered pressure distribution will be presented by fixing one parameter and varying the other. Note that the scattered pressure mentioned here is due to wall pressure gusts with $k_2 = 0$. The discussion, however, also applies to gusts with $k_2 \neq 0$, as the streamwise number $k_1$ has the same value for different gusts.

The scattered pressure on the flat plate is presented in figure~\ref{fig:surfacePressureSerratedFreq} for different values of $k_1 h$. The scattered surface pressure is obtained by evaluating the real part of (\ref{equ:ScatteredSurfacePressure1st}) using the second-order approximation ($P_i = 1$) and then normalizing to unity. The results presented in figure~\ref{fig:surfacePressureSerratedFreq} are obtained for $k_1 h_e = 7$, while $k_1 h$ varies between $2$ and $20$. The horizontal coordinate~$y^\prime$ is normalized by the spanwise correlation length $l_{y^\prime}$ and the vertical coordinate~$x^\prime$ is normalized by the hydrodynamic wavelength $\lambda_1 = 2\pi/k_1$. Thus, the distance between the two adjacent vertical dashed lines corresponds to the spanwise correlation length, $l_{y^\prime}$. Figure~\ref{fig:surfacePressureSerratedFreq}(a) plots the scattered surface pressure distribution near the trailing-edge for $k_1 h = 2$. It shows that the scattered surface pressure field between two adjacent vertical lines is essentially in phase, so no strong phase variation within $l_{y^\prime}$ occurs. Figure~\ref{fig:surfacePressureSerratedFreq}(b) shows the scattered surface pressure distribution for $k_1 h = 4$, and it can seen that little phase differences appear within a spanwise correlation length. Figure~\ref{fig:surfacePressureSerratedFreq}(c) shows the scattered surface pressure distribution for $k_1 h = 10$. It is clear that even if the spanwise correlation length becomes smaller, a pronounced phase difference still appears within adjacent vertical lines. Further increasing the value of $k_1 h$ to $20$, as shown in figure~\ref{fig:surfacePressureSerratedFreq}(d), decreases the spanwise correlation length, but enough phase difference still appears within the increasingly narrow ranges.

To make the phase variation induced by the presence of serrations even clearer, the scattered surface pressure along the trailing-edge is presented in figure~\ref{fig:pressureOnEdgeFrequency}. Each line corresponds to a different value of $k_1h$. The real and imaginary parts of the pressure are shown in figure~\ref{fig:pressureOnEdgeFrequency}(a) and \ref{fig:pressureOnEdgeFrequency}(b), respectively. The two figures are thus showing the pressure distributions at different instants. The red curve in figure~\ref{fig:pressureOnEdgeFrequency}(a), which corresponds to the real part for $k_1 h = 2$, remains almost entirely negative. The corresponding imaginary part, shown in red in figure~\ref{fig:pressureOnEdgeFrequency}(b), has a phase which slightly changes signs over $l_{y^\prime}$. Since the signal oscillates between the real and imaginary parts, the phase chagnes sign only over a small fraction of the cycle. The black curves corresponding to $k_1h =20$, on the other hand, show a strong variation within a spanwise correlation length in both figure~\ref{fig:pressureOnEdgeFrequency}(a) and \ref{fig:pressureOnEdgeFrequency}(b), indicating a strong phase variation over the whole cycle. Therefore, the phase differences of the scattered pressure are more likely to be strong and permanent for high values of $k_1 h$.


The scattered pressure distributions for different values of $k_1 h_e$ are presented in figure~\ref{fig:surfacePressureSerratedSharpness}. The values of $k_1 h$ is fixed at $10$ while $k_1 h_e$ increases from $1$ to $10$ (see figures~\ref{fig:surfacePressureSerratedSharpness}(a) to \ref{fig:surfacePressureSerratedSharpness}(d)). The horizontal and vertical coordinates are also normalized by $l_{y^\prime}$ and $\lambda_1$, respectively. Figure~\ref{fig:surfacePressureSerratedSharpness}(a) presents the scattered surface pressure distribution for $k_1 h_e = 1$. As $k_1 h_e$ is small, the distance between two adjacent dashed lines is very small compared to the serration wavelength. Thus, even though pronounced phase differences appear along the edge, one can hardly see any phase variations within a spanwise correlation length. The pressure distribution for $k_1 h_e = 2$ is shown in figure~\ref{fig:surfacePressureSerratedSharpness}(b) and no significant phase variations are achieved. However, for $k_1 h_e = 4$ a clear phase variation of scattered pressure begins to appear within adjacent lines, as shown in figure~\ref{fig:surfacePressureSerratedSharpness}(c). From figure~\ref{fig:surfacePressureSerratedSharpness}(d), where $k_1 h_3 = 10$, it can be seen that pronounced phase differences appear within a correlation length $l_{y^\prime}$ in the spanwise direction. To better visualize the phase variation appearing in the spanwise direction due to serrations, the scattered pressure along the serrated edge is shown in figure~\ref{fig:pressureOnEdgeSharpness} for the same values of $k_1 h_e$ as in figure~\ref{fig:surfacePressureSerratedSharpness}. Both the real and imaginary parts of the scattered pressure are presented. The tendency for large values of $k_1 h_e$ to favor strong phase variations is clearly demonstrated.

Finally, one is in a position to discuss the noise reduction mechanism by investigating the physical implications of the two parameters $k_1h$ and $k_1 h_e$. From figure~\ref{fig:surfacePressureSerratedFreq} it is obvious that the criterion $k_1 h \gg 1$ ensures an effective phase variation appearing along and near the trailing-edge in the spanwise direction. In addition, as shown from figure~\ref{fig:surfacePressureSerratedSharpness}, the condition $k_1 h_e \gg 1$ ensures that the phase difference appears within one spanwise correlation length, i.e. between two adjacent dashed lines shown in figure~\ref{fig:surfacePressureSerratedFreq} and \ref{fig:surfacePressureSerratedSharpness}. Since the surface pressure within the regions bordered by dashed lines is coherent, and since the range over which the phase difference appears is clearly much smaller than the acoustic wavelength, the far-field sound will be reduced due to destructive interference. Physically, this means that the phase differences induced on the flat plate in the spanwise direction due to the presence of serrations should be well situated within a correlated turbulent structure. Therefore, as demonstrated in both figures~\ref{fig:surfacePressureSerratedFreq} and \ref{fig:surfacePressureSerratedSharpness}, the sound reduction is caused by the destructive interference of the scattered surface pressure due to the presence of serrations.

Results in Sec.~\ref{subsec:Results} have shown that in the case of wide serrations, a noise increase at low frequencies is also possible, especially at high Mach numbers. The reason is that at low frequencies, only little phase variation is induced by the presence of serrated trailing-edges in the spanwise direction, but the wetted length of the trailing-edge is in fact much longer than that of a straight one. Thus, the net effect of phase interference can be constructive, which leads to an increase in the far-field noise. This is more likely to occur when $k_1 h$ is small, i.e. $M_0$ is large, frequency is low or the serration is wide (small value of $h$).

%Otherwise, the phase differences induced are not ``real'' phase differences, in the sense that they are not ``fully correlated''. This can be made clear when one considers a limit case, e.g. when the spanwise correlation length is zero: the scattered pressure on every segment of an edge is statistically independent, so no cancellation is possible.

Based on the preceding discussions, one can conclude that the minimum effective serration length required for noise reduction can be obtained from $h_{min} = \min(h, h_e)$. It is then straightforward to combine the two conditions stated above into one, i.e. $k_1 h_{min} \gg 1$. Therefore, in order to achieve an effective noise reduction in the far-filed, the geometry of the serrations should satisfy $k_1 h_{min} \gg 1$. Based on the interference results in figure~\ref{fig:surfacePressureSerratedFreq} and \ref{fig:surfacePressureSerratedSharpness}, it can be found that a common rule of thumb is $k_1 h_{min} \gtrsim \pi$, with higher value more favourable.

\section{Conclusion}
\label{sec:VI}
A new mathematical model is developed in this paper to predict the sound radiated by serrated trailing-edges. The model begins with establishing an idealized scattering problem, resulting in a mixed boundary value convective wave equation problem with complex boundaries. This leads to a set of coupled partial differential equations, which cannot be solved using the standard separation method. A solution is obtained based on Fourier expansion to separate the variables and Schwarzschild's method together with an iterative technique to solve the resulting coupled equations. The far-field sound is evaluated using the surface pressure integrals. The PSD of far-field sound is related to the wavenumber spectral density of the wall pressure beneath the turbulent boundary layer near the trailing-edge using Amiet's approach~\citep{Amiet1976, Amiet1978}.

The results obtained using the new model agree well with FEM computations, suggesting that the model developed in this paper captures the scattering process and gives correct predictions for the sound generated by serrated trailing-edges. It is shown that the coupling effect must not be ignored and as a result the new model can predict the sound reduction more accurately than Howe's model~\citep{Howe1991,Howe1991a}. The results obtained using the new model agree better with experiments, in which the average sound reduction is reported to be up to around 7 dB. The directivity results show that serrations can significantly reduce the noise in the area near the leading edge and that at high Mach numbers the use of serrations can lead to noise increase at small angles.

The physical mechanism for noise reduction is found to be interference effects in the wall pressure fluctuations due to the presence of serrations. Two non-dimensional parameters are found to be critical. First, $k_1h \gg 1$ to ensure the existence of strong phase variation in the spanwise direction. Second, $k_1h_e \gg 1$ to ensure that the phase differences along the edges are correlated in the spanwise direction. The sound reduction generally increases as the serration sharpness increases, but if the serrations are already sharp enough, further increasing the slope only affects high frequencies.

The results obtained using Chase's turbulent boundary layer spectrum model do not appear to explain the noise increase observed in experiments at high frequencies~\citep{Parchen1999, Oerlemans2009, Gruber2012}. This suggests that the wavenumber-frequency spectra of the surface pressure fluctuations (see Sec.~\ref{subsec:Discussion}) are not accurate or that this noise increase is due to some other mechanisms, such as the high intensity flow through serration valleys.
In addition, the current model assumes perfect correlations in the streamwise direction, which may not be sufficiently accurate according to the measurements of~\cite{Gruber2012}. Thus, the current model may be further improved by incorporating more physical parameters such as the streamwise correlation length.
% Another possibility is that the streamwise correlations play an % important role and should be incorporated into the trailing edge noise % model for serrated edges.

\section*{Acknowledgments}
The first author (BL) wishes to gratefully acknowledge the financial support co-funded by the Cambridge Commonwealth European and International Trust and China Scholarship Council. The second author (MA) would like to acknowledge the financial support of the Royal Academy of Engineering. The third author (SS) wishes to gratefully acknowledge the support of the Royal Commission for the exhibition of 1851. Finally, the authors thank Stephane Moreau and Michel Roger for their useful feedback and many stimulating discussions, and Professor Dame Ann Dowling for helping to resolve an early error on modelling the scattered pressure jump.

\appendix
\section{}{\label{AppendixA}}
\subsection{Scattered pressure of second iteration}
The solution after the second iteration can be expressed as,
\begin{equation}
	\boldsymbol{P}^{(2)}(x,0) = \boldsymbol{N}(x) + \boldsymbol{C}^{(1)}(x) + \boldsymbol{C}^{(2)}(x),
	\label{secondsolution}
\end{equation}
where $\boldsymbol{N}(x)$ and $\boldsymbol{C}^{(1)}(x)$ are defined in Sec.~\ref{sec:II}, and $\boldsymbol{C}^{(2)}(x)$ whose entry corresponding to mode $n^{\prime}$ is
\begin{equation}
	\begin{aligned}
            C_{n^{\prime}}^{(2)}(x)= & P_{i}(1-i)\mathrm{e}^{\mathrm{i}k_1x} \sum_{m=-\infty}^\infty\Bigg\{ \beta_{n^{\prime}m} (\mathrm{i}k_1)^2 (\mathrm{E}(-\mu_{n^{\prime}}x)- \mathrm{E}(-\mu_mx))\\
            & - \big(\beta_{n^{\prime}m}\mathrm{i}k_1 + \gamma_{n^{\prime}m} \mathrm{i}(k_1-\mu_m)\big)\sqrt{\frac{\mu_m}{-2\pi x}} (\mathrm{e}^{-\mathrm{i}\mu_{n^{\prime}}x}- \mathrm{e}^{-\mathrm{i}\mu_mx})  \\
            & -\frac{\gamma_{n^{\prime}m}}{2}\big(\sqrt{\frac{\mu_m}{-2\pi x}}\frac{1}{(-x)}(\mathrm{e}^{-\mathrm{i}\mu_{n^{\prime}}x}- \mathrm{e}^{-\mathrm{i}\mu_mx})  - \mathrm{i}(\mu_{n^{\prime}}-\mu_m)\sqrt{\frac{\mu_m}{-2\pi x}}\mathrm{e}^{-\mathrm{i}\mu_{n^{\prime}}x}\big)\Bigg\},
\end{aligned}
\label{equ:secondOrderPart}
\end{equation}
where
\begin{equation*}
\begin{aligned}
&\beta_{ln} = \sum_{m=-\infty}^{\infty} \big(v_{ln}a_m - B_{lm}/(k_{2l}^2-k_{2n}^2)a_n \big) v_{nm},\\
&\gamma_{ln} = \sum_{m=-\infty}^{\infty}\big(v_{ln}a_m \sqrt{\mu_m/\mu_n} - B_{lm}/(k_{2l}^2-k_{2n}^2)a_n \big)v_{nm}.\\
\end{aligned}
\end{equation*}
\nomenclature[g-beta]{$\beta_{ln}$}{Coefficient denoting higher order coupling}
\nomenclature[g-gamma]{$\gamma_{ln}$}{Coefficient denoting higher order coupling}

\subsection{Far-field sound pressure of second iteration}
The function $T_{nm}$ involved in the second iteration can be defined as:
\begin{equation}
	\begin{aligned}
            T_{nm} = &\sum_{j=0}^1 \frac{1}{\mathrm{i}\kappa_{nj}}\Bigg\{\Big( \frac{\mathrm{i}\eta_{Am}}{\sqrt{\eta_{Am}}} \big(\mathrm{e}^{\mathrm{i}\kappa_{nj} \chi_{j+1}}\mathrm{E}(\eta_{Am}(c+\epsilon_{j+1})) - \mathrm{e}^{\mathrm{i}\kappa_{nj} \chi_{j}}\mathrm{E}(\eta_{Am}(c+\epsilon_{j}))\big)\\
            &  - \frac{\mathrm{i}\eta_{Bmj}}{\sqrt{\eta_{Bmj}}} \mathrm{e}^{\mathrm{i}\kappa_{nj} (\chi_j - (c+\epsilon_j)/\sigma_j)}\big(\mathrm{E}(\eta_{Bmj}(c+\epsilon_{j+1})) - \mathrm{E}(\eta_{Bmj}(c+\epsilon_{j}))\big) \Big) -\\
            \Big(& \big(\mathrm{e}^{\mathrm{i}\kappa_{nj} \chi_{j+1}}\frac{1}{\sqrt{2\pi(c+\epsilon_{j+1})}}\mathrm{e}^{\mathrm{i}\eta_{Am}(c+\epsilon_{j+1})} - \mathrm{e}^{\mathrm{i}\kappa_{nj} \chi_{j}}\frac{1}{\sqrt{2\pi(c+\epsilon_{j})}}\mathrm{e}^{i\eta_{Am}(c+\epsilon_{j})}\big)\\
            &  -  \mathrm{e}^{\mathrm{i}\kappa_{nj} (\chi_j - (c+\epsilon_j)/\sigma_j)}\big(\frac{1}{\sqrt{2\pi(c+\epsilon_{j+1})}}\mathrm{e}^{\mathrm{i}\eta_{Bmj}(c+\epsilon_{j+1})} - \frac{1}{\sqrt{2\pi(c+\epsilon_{j})}}\mathrm{e}^{i\eta_{Bmj}(c+\epsilon_{j})}\big)\Big)\Bigg\},
	\end{aligned}
	\label{S_nm^2}
\end{equation}

The second iterated solution falls into the same pattern,
\begin{equation}
	\begin{aligned}
            p^{(2)}(\mathbf{x},\omega) = &\frac{-\mathrm{i}\omega x_3}{2\pi c_0 S_0^2} P_{i} \mathrm{e}^{-\mathrm{i}k/\beta^2(Mx_1-S_0)} \mathrm{e}^{\mathrm{i}k/\beta^2(M-x_1/S_0)h}(1-\mathrm{i})  \times \\
	&\frac{\sin\big((N+1/2)\lambda (k_2-kx_2/S_0)\big)}{\sin\big(\lambda/2(k_2-kx_2/S_0)\big)}\sum_{n^{\prime}=-\infty}^\infty \Bigg(\Theta_{n^{\prime}} + \Theta_{n^{\prime}}^{(1)} + \Theta_{n^{\prime}}^{(2)}  \Bigg),
	\label{2ndsound}
\end{aligned}
\end{equation}
where $\Theta_{n^{\prime}}$ and $\Theta_{n^{\prime}}^{(1)}$ are defined in Sec.~\ref{sec:II}, and
\begin{equation}
	\begin{aligned}
            \Theta_{n^{\prime}}^{(2)} =& \sum_{m=-\infty}^\infty \beta_{n^{\prime}m}(\mathrm{i}k_1)^2(Q_{n^{\prime}n^{\prime}} - Q_{n^{\prime}m}) - (\beta_{n^{\prime}m}\sqrt{\mu_{m}} \mathrm{i}k_1 + \gamma_{n^{\prime}m}\sqrt{\mu_m}\mathrm{i}(k_1-\mu_m))(S_{n^{\prime}n^{\prime}} - S_{n^{\prime}m} ) \\
            & - \gamma_{n^{\prime}m}\sqrt{\mu_m}\big( T_{n^{\prime}n^{\prime}} - T_{n^{\prime}m} -\mathrm{i}/2(\mu_{n^{\prime}}-\mu_m)S_{n^{\prime}n^{\prime}}\big).
\end{aligned}
\end{equation}

% Bibliography

\input{references.bbl}
\end{document}

%% file: jfmdefs.tex
\ifCUPmtlplainloaded \else
  \checkfont{eurm10}
  \iffontfound
    \IfFileExists{upmath.sty}
      {\typeout{^^JFound AMS Euler Roman fonts on the system,
                   using the 'upmath' package.^^J}%
       \usepackage{upmath}}
      {\typeout{^^JFound AMS Euler Roman fonts on the system, but you
                   dont seem to have the}%
       \typeout{'upmath' package installed. JFM.cls can take advantage
                 of these fonts,^^Jif you use 'upmath' package.^^J}%
      }
  \else
  \fi
\fi

\ifCUPmtlplainloaded \else
  \checkfont{msam10}
  \iffontfound
    \IfFileExists{amssymb.sty}
      {\typeout{^^JFound AMS Symbol fonts on the system, using the
                'amssymb' package.^^J}%
       \usepackage{amssymb}%
       \let\le=\leqslant  
       \let\ge=\geqslant  \let\geq=\geqslant
      }{}
  \fi
\fi

\ifCUPmtlplainloaded \else
  \IfFileExists{amsbsy.sty}
    {\typeout{^^JFound the 'amsbsy' package on the system, using it.^^J}%
     \usepackage{amsbsy}}
    {\providecommand\boldsymbol[1]{\mbox{\boldmath $##1$}}}
\fi

\newsavebox{\astrutbox}
\sbox{\astrutbox}{\rule[-5pt]{0pt}{20pt}}